\documentclass[twocolumn, trackchanges]{aastex63}

\usepackage{graphicx, amsmath}
\usepackage[caption=false]{subfig}
\usepackage{hyperref}
\usepackage{cleveref}
\usepackage{xspace}
\usepackage{rotating}
\usepackage{comment}

\newcommand{\hetdex}{HETDEX}
\newcommand{\lya}{Ly$\alpha$\xspace}
\newcommand{\nsample}{935}
\newcommand{\Ha}{H$\alpha$\xspace}
\newcommand{\Hb}{H$\beta$\xspace}

\newcommand{\OII}{[\ion{O}{2}]}
\newcommand{\OIII}{[\ion{O}{3}]}

\newcommand{\NII}{[\ion{N}{2}]}

\newcommand{\mcsed}{\texttt{MCSED}}

\newcommand{\threedhst}{\mbox{3D-HST}}

\newcommand{\fesc}{$f_{\rm esc}^{\rm Ly\alpha}$\xspace}

\def\ecs{{ergs~cm$^{-2}$~s$^{-1}$}}
\def\etal{{et~al.\null}}

\graphicspath{{./}{figures/}}

\begin{document}

\title{The \hetdex\ Survey: The \lya Escape Fraction from \threedhst\ Emission Line Galaxies at $\lowercase{z}\sim2$ }

\author{Laurel H. Weiss}
\affiliation{Department of Astronomy \& Astrophysics, The Pennsylvania State University, University Park, PA 16802}
\email{laurelhweiss@gmail.com}

\author[0000-0003-4381-5245]{William P. Bowman}
\affiliation{Department of Astronomy \& Astrophysics, The Pennsylvania
State University, University Park, PA 16802}
\affiliation{Institute for Gravitation and the Cosmos, The Pennsylvania State University, University Park, PA 16802}

\author[0000-0002-1328-0211]{Robin Ciardullo}
\affiliation{Department of Astronomy \& Astrophysics, The Pennsylvania
State University, University Park, PA 16802}
\affiliation{Institute for Gravitation and the Cosmos, The Pennsylvania State University, University Park, PA 16802}

\author{Gregory R. Zeimann}
\affiliation{Hobby Eberly Telescope, University of Texas, Austin, Austin, TX, 78712, USA}

\author{Caryl Gronwall}
\affiliation{Department of Astronomy \& Astrophysics, The Pennsylvania
State University, University Park, PA 16802}
\affiliation{Institute for Gravitation and the Cosmos, The Pennsylvania State University, University Park, PA 16802}

\author{Erin Mentuch Cooper}
\affiliation{Department of Astronomy, The University of Texas at Austin, 2515 Speedway Boulevard, Austin, TX 78712, USA}

\author{Karl Gebhardt}
\affiliation{Department of Astronomy, The University of Texas at Austin, 2515 Speedway Boulevard, Austin, TX 78712, USA}

\author{Gary J. Hill}
\affiliation{Department of Astronomy, The University of Texas at Austin, 2515 Speedway Boulevard, Austin, TX 78712, USA}
\affiliation{McDonald Observatory, The University of Texas at Austin, 2515 Speedway, Austin, TX 78712, USA}

\author{Guillermo A. Blanc}
\affiliation{Departamento de Astronom\'ia, Universidad de Chile, Casilla 36-D,
Santiago, Chile}
\affiliation{The Observatories of the Carnegie Institution for Science, 813
Santa Barbara Street, Pasadena, CA 91101, USA}

\author{Daniel J. Farrow}
\affiliation{Max-Planck-Institut f{\"u}r extraterrestrische Physik, Postfach 1312 Giessenbachstrasse, 85741 Garching, Germany}

\author[0000-0001-8519-1130]{Steven L. Finkelstein}
\affiliation{Department of Astronomy, The University of Texas at Austin, 2515 Speedway Boulevard, Austin, TX 78712, USA}

\author{Eric Gawiser}
\affiliation{Department of Physics and Astronomy, Rutgers, The State University of New Jersey, 136 Frelinghuysen Road, Piscataway, NJ 08854, USA}

\author{Steven Janowiecki}
\affiliation{McDonald Observatory, The University of Texas at Austin, 2515 Speedway, Austin, TX 78712, USA}

\author{Shardha Jogee}
\affiliation{Department of Astronomy, The University of Texas at Austin, 2515 Speedway Boulevard, Austin, TX 78712, USA}

\author{Donald P. Schneider}
\affiliation{Department of Astronomy \& Astrophysics, The Pennsylvania
State University, University Park, PA 16802}
\affiliation{Institute for Gravitation and the Cosmos, The Pennsylvania State University, University Park, PA 16802}

\author{Lutz Wisotzki}
\affiliation{Leibniz-Institut f\"ur Astrophysik Potsdam (AIP), An der Sternwarte 16, 14482 Potsdam, Germany}

\begin{abstract}
We measure the \lya escape fraction of 935 \OIII-emitting galaxies between $1.9 < z < 2.35$ by comparing stacked spectra from the \textit{Hubble Space Telescope}/WFC3's near-IR grism to corresponding stacks from the Hobby Eberly Telescope Dark Energy Experiment's Internal Data Release 2. By measuring the stacks' \Hb to \lya ratios, we determine the \lya escape fraction as a function of stellar mass, star formation rate, internal reddening, size, and \OIII/\Hb ratio.  We show that the escape fraction of \lya correlates with a number of parameters, such as galaxy size, star formation rate, and nebular excitation. However, we also demonstrate that most of these relations are indirect, and the primary  variables that control the escape of \lya are likely stellar mass and internal extinction.   Overall, the escape of \lya declines from $\gtrsim 18\%$ in galaxies with $\log M/M_{\odot} \lesssim 9$ to $\lesssim 1\%$ for systems with $\log M/M_{\odot} \gtrsim 10$, with the sample's mean escape fraction being $6.0^{+0.6\%}_{-0.5\%}$.
 
\end{abstract}

\keywords{cosmology: observations -- galaxies: evolution -- galaxies: formation -- galaxies: high redshift}


\section{Introduction}\label{sec:intro}
Lyman alpha (\lya) is a strong indicator of star formation and has long been regarded as one of the best probes for identifying galaxies in the act of formation \citep{Partridge_1967}. High-energy photons from young, hot stars ionize hydrogen and produce electrons, which later recombine with hydrogen nuclei and cascade down through the energy levels. These electrons are eventually funneled into the $n=2$ state by the large cross section for Lyman line absorption in the interstellar medium (ISM\null).  Roughly 68\% of these $n=2$ electrons decay to create Ly$\alpha$, which can then be resonantly scattered throughout interstellar space. 

The fate of these \lya photons is highly sensitive to the geometry and physical conditions of the interstellar medium; even trace amounts of dust can break the chain of interactions needed for \lya to escape into intergalactic space  \citep[e.g.,][]{Ahn_2001, Verhamme_2006, Dijkstra_2006, Rivera-Thorsen_2015}.  As a result, \lya measurements not only reflect the initial ionization from hot young stars, but also the total amount of dust in the interstellar medium \citep[e.g.,][]{Finkelstein_2009, Schaerer_2011}, the distribution of the dust \citep[e.g.,][]{Finkelstein_2008, Scarlata_2009}, the ISM kinematics \citep[e.g.,][]{Kunth_1998}, and the galaxy's neutral gas content 
and gas geometry  \citep[e.g.,][]{Neufeld_1991, Hansen_2006, Jaskot_2014}.  Observations of \lya escape at different cosmic epochs therefore provide a unique tool with which to probe the evolution of the interstellar medium within ensembles of galaxies.  For recent reviews of the subject, see \citet{Dijkstra_2014} and \citet{Hayes_2015}.

Several studies have attempted to indirectly measure the fraction of \lya escaping a galaxy (\fesc) using various measures of a galaxy's star formation rate (SFR\null).  Observations in the rest-frame ultraviolet (UV), X-ray, and infrared (IR) can all be translated into SFRs via local calibrations \citep[e.g.,][and references therein]{Kennicutt_2012}.  That rate, in turn, can be used to predict the production rate of Ly$\alpha$, which can then be compared to  observations \citep[e.g.,][]{Gronwall_2007, Blanc_2011, Steidel_2011, Zheng_2012, Wardlow_2014, Oyarzun_2017}.  Obviously, analyses such as these have their limitations, as the inferred escape fractions are only as accurate as the SFR estimates and the assumed behavior of the initial mass function.  Moreover, while a galaxy's emission lines, UV continuum, far-IR emission, and X-ray flux all indicate the presence of hot, young stars, emission lines such as \lya originate from ionizing photons produced in the most massive stars ($M \gtrsim 15 M_{\odot}$) and therefore only reflect star formation over the past $t \lesssim 10$~Myr.  In contrast, other SFR indicators probe longer ($t \sim 100$~Myr) timescales and are sensitive to more of the initial mass function \citep[$M \gtrsim 5 M_{\odot}$;][]{Kennicutt_2012}. Thus comparisons which use these SFR indicators are susceptible to increased errors, which, due to the evolving SFR density of the universe, may be systematic.

The most direct method of measuring the escape fraction of \lya is to compare a system's \lya flux to its Balmer emission.  In steady state, the rate of decays into hydrogen's $n=2$ state must equal the number of decays out of that state.  Since the intrinsic Balmer line ratios depend very weakly upon the conditions in the ISM, the ratio of \lya to a single line of hydrogen, such as \Ha or \Hb, yields a robust measurement of the escape fraction of \lya.

\cite{Hayes_2010} were the first to attempt such a measurement using deep, narrowband images for \lya and \Ha at $z \sim 2.2$.  By comparing the resultant luminosity functions, \citet{Hayes_2010} estimated the epoch's volumetric \lya escape fraction to be $5.3 \pm 3.8\%$. However, this estimate was limited by the survey's small volume and lack of sensitivity; in particular,  the survey only covered $\sim 5440$~Mpc$^3$ and detected just 55 galaxies in H$\alpha$.  \citet{Ciardullo_2014} confirmed the \citet{Hayes_2010} measurement by comparing \Hb detections in a subset of 73 $1.90 < z < 2.35$ galaxies identified in the \threedhst\ IR-grism program \citep{Brammer_2012} to \lya data from integral-field unit (IFU) spectroscopy from the Hobby Eberly Telescope Dark Energy Experiment Pilot Survey \citep{Adams_2011, Blanc_2011}. The authors reported a volumetric \lya escape fraction of $4.4^{+2.1}_{-1.9}\%$, although this study was again limited by a small sample size.

Neither of these early analyses examined trends in the behavior of \lya emission with the physical properties of the galaxies.  To address this question, several studies have attempted to estimate the escape of \lya in various sets of both continuum and emission-line selected galaxies. Optical and infrared spectroscopy on a sample of color-selected $2 \lesssim z \lesssim 3$ systems found that Ly$\alpha$ emission is more common in older, relatively dust-free galaxies \citep{Kornei_2010} and systems with extreme \NII/H$\alpha$ and \OIII/H$\beta$ line ratios \citep{Erb_2016}.  Moreover,  \cite{Oyarzun_2016} and \citet{Oyarzun_2017} concluded that the amount of \lya that escapes a galaxy is anti-correlated with its stellar mass over the range $8 \lesssim \log (M/M_{\odot}) \lesssim 10.5$. \citet{Hagen_2016}, however, detected no such correlation in their comparison of the physical properties of $z \sim 2$ \OIII\ and \lya emitters, while \citet{Shimakawa_2017} argued that \lya emitting galaxies are similar to normal star-forming systems in the low-mass regime, but are a unique population above $\sim 10^{10} ~M_{\odot}$.  

All of the above results rely on indirect measurements of the \lya escape fraction, are based on small samples sizes, or are compromised in some way by selection effects.  In this paper, we mitigate these issues by examining the \lya emission from a sample of \nsample\ galaxies between $1.90 < z < 2.35$ selected on the basis of their rest-frame optical emission lines as measured by the G141 grism of the \textit{Hubble Space Telescope's} Wide Field Camera 3\null. Specifically, we compare \lya measurements from the Hobby Eberly Telescope Dark Energy Experiment (HETDEX; \citealt{Hill_2008, hil16}; Gebhardt \etal\ 2021, in prep.) with \Hb detections from the \threedhst\ survey \citep{Brammer_2012, Momcheva_2016} and measure the escape fraction of \lya as a function of stellar mass, galaxy size, star formation rate, galactic internal extinction, and \mbox{\OIII\ $\lambda 5007$/\Hb} ratio. Section~\ref{sec:sample} describes the observational data, as well as the physical properties of the sample.  Section~\ref{sec:HETDEX} summarizes the instrumentation of the HETDEX project and states how the \lya fluxes were obtained from the VIRUS spectra.  Section~\ref{sec:fesc} reviews how line fluxes from \threedhst\ and HETDEX can be leveraged to estimate the \lya escape fraction, and details how we correct our measurements for the attenuation of \Hb in the rest-frame optical spectra.  Section~\ref{sec:results} presents our results:  Section~\ref{subsec:singleobjects} summarizes the detections in individual galaxies, Section~\ref{subsec:stacking} describes our methodology for stacking spectra, Section~\ref{subsec:stacking_results} uses our stacks to measure how the \lya escape fraction changes with stellar mass, SFR, reddening, size, and emission-line ratio, and Section~\ref{subsec:guang} attempts to disentangle the dependencies to identify those properties which are most important for the escape of \lya.  We conclude by estimating the mean escape fraction for the sample and discussing our results in the context of other studies of \lya in the $z \gtrsim 2$ universe.

We assume a $\Lambda$CDM cosmology, with $\Omega_{\Lambda}$= 0.7,  $\Omega_{M}$= 0.3 and $H_0$ = 70 $\mathrm{km~s^{-1}~Mpc^{-1}}$. 

\section{The Sample}
\label{sec:sample}
This study uses a sub-sample of 935 $z \sim 2$ emission-line galaxies (ELGs) originally identified in the AEGIS \citep{Davis_2007}, COSMOS \citep{Scoville_2007}, and GOODS-N \citep{Giavalisco_2004} fields by the \threedhst\ grism survey \citep{Brammer_2012, Momcheva_2016}.  These galaxies are part of a carefully vetted set of objects selected by \citet{Bowman_2019}, all with apparent $JH$ magnitudes $\leq 26$, unambiguous emission-line-based redshifts between $1.90 \leq z \leq 2.35$ (the range where both \OII\ $\lambda 3727$ and \OIII\ $\lambda 5007$ are visible), and a 50\% line-flux completeness limit of $\sim 4 \times 10^{-17}$~ergs~cm$^{-2}$~s$^{-1}$.  In this redshift range, the co-moving volume under study is $\sim 5 \times 10^5$~Mpc$^3$.

\begin{figure*}
  \centering
  \subfloat[]{%
    \includegraphics[height=7.25cm]{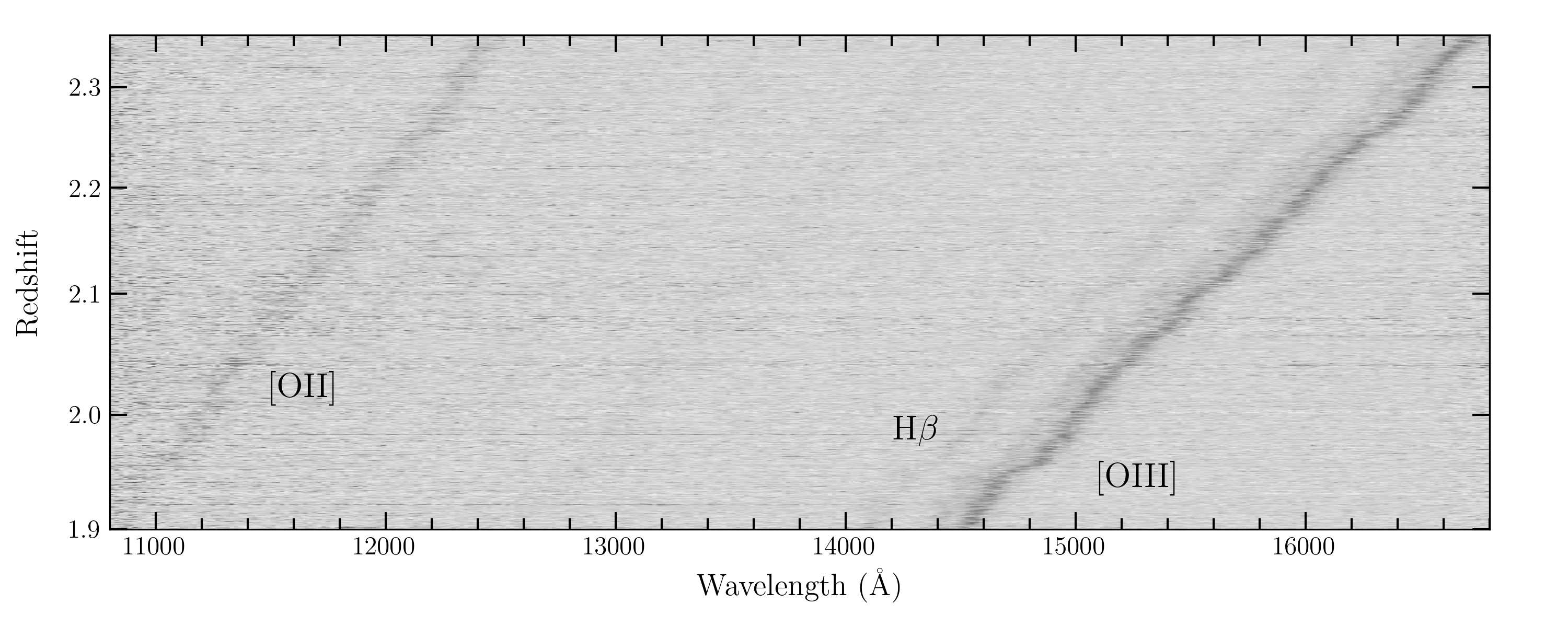}%
  }
  \caption{Redshift-ordered \threedhst\ grism spectra for \nsample\ emission-line selected galaxies.  The spectra are continuum-subtracted and normalized by their \OIII\ flux. The brightest feature is the distinctively-shaped blend of \OIII\ $\lambda 4959+\lambda 5007$.  The data products are taken from the \threedhst\ catalog \citep{Brammer_2012, Momcheva_2016}.} 
  \label{fig:grism}
\end{figure*}

Figure~\ref{fig:grism} displays these continuum-subtracted grism spectra in ascending order of redshift, normalized by their \OIII\ flux.  In over 90\% of the sample, the \OIII~doublet is the strongest feature; in 90\% of the remaining galaxies, \OII\ dominates.  Most AGN have been removed from this dataset via comparisons with X-ray source catalogs; \citet{Bowman_2019} estimate the fraction of remaining AGN in the sample to be less than 5\%. 

\textit{HST} images of our galaxies in the rest-frame UV and optical are presented in \citet{Koekemoer_2011}, the galaxies' line fluxes come from the catalog of \citet{Momcheva_2016}, and multiwavelength PSF-matched photometry is given by \citet{Skelton_2014}.  \citet{Bowman_2019} used these data, along with the \mcsed\ spectral energy distribution (SED)-fitting code to estimate the galaxies' 
half-light radii, stellar masses, star-formation rates, and dust attenuation.  Improved values for the latter three quantities, based on more realistic fitting assumptions such as a four age-bin star formation history (with borders at [0.001, 0.1, 0.3, 1.0, 3.2 Gyr]) and a three-parameter dust attenuation law \citep{nol09}, are given in \citet{Bowman_2020}.  These fits, which use the spectral libraries of the Flexible Stellar Population Synthesis code \citep{fsps-1, fsps-2} and the grids of nebular emission by \citet{byler+17}, assume PADOVA isochrones \citep{bertelli+94, girardi+00, marigo+08}, a \citet{cha03} initial mass function, an ionization parameter consistent with the observed line ratios ($\log U = -2.5$), and treat metallicity as a free parameter.  The fits show that the galaxies in our sample have stellar masses between $8.0 \lesssim \log M/M_{\odot} \lesssim 10.5$, SFRs between $1 \lesssim \textrm{SFR} \lesssim 100 \,  M_{\odot}$~yr$^{-1}$, internal extinctions between $0.0 \lesssim E(B-V) \lesssim 0.8$, and optical half-light radii between $0.5 \lesssim R_e \lesssim 5$~kpc.

\section{Optical Spectroscopy}
\label{sec:HETDEX}
Of the 1952 galaxies in the \citet{Bowman_2019} sample, 1210 are located in the AEGIS, COSMOS, and GOODS-N fields.  As of 1-July-2020, \nsample\ of these objects had been successfully observed at least once (and most multiple times) with the Visible Integral-Field Replicable Unit Spectrographs (VIRUS, \citealt{hil18a}) of the upgraded Hobby Eberly Telescope (HET, \citealt{hil18b}) as part of HETDEX science verification.  VIRUS consists of an array of up to 78 IFU spectrographs, each covering an area of $51\arcsec \times 51\arcsec$ on the sky.  Each of these IFUs  contains a set of 448 $1\farcs 5$-diameter optical fibers arranged in a hexagonal close-pack pattern, which feeds a pair of low-resolution ($R \sim 800$) spectrographs covering the wavelength range between 3500~\AA\ and 5500~\AA\ at a dispersion of $\sim 2$~\AA\ per (2$\times$ binned) pixel.  The result is an instrument system capable of obtaining 34,944 spectra per exposure (Hill \etal\ 2021, in prep).

VIRUS does not have lenslets, and the center of each $1\farcs 5$ diameter fiber is offset from its nearest neighbor by $2\farcs 543$.  This results in an IFU fill-factor of 1/3 per single exposure. To fill in the gaps between fibers, each HETDEX observation consists of three 6-min exposures (scaled for observing conditions) taken in a triangular dither pattern.  As the typical seeing at the HET is slightly greater than the VIRUS fiber size, this dithering effectively produces a uniform fill factor within the IFUs and helps sample the point spread function of the observation.   Moreover, since the HETDEX science verification fields have been visited multiple times during the project's first 2.5 years, most of our targets have several measurements, with the deepest fields targeted 21 times.  Table~\ref{tab:observation} summarizes the number of observations per object for our \nsample\ galaxies.  Only those observations suitable for analysis are listed in the table. 

\begin{deluxetable}{lcc}
\tablecaption{\hetdex\ Observations of $1.90 \leq z \leq 2.35$ \OIII\ emitters \label{tab:observation}}
\tablewidth{0 pt}
\tablehead{
\colhead{$N_{\rm observations}$} & \colhead{$N_{\rm objects}$}}
\startdata
 1 & 294 \\
 $2 \leq N \leq 3$  & 324 \\
 $4 \leq N \leq 6$  & 217 \\
 $7 \leq N \leq 10$ & 77 \\
 $N \geq 11$        & 23 
\enddata
\end{deluxetable}

The data processing of the \hetdex\ frames is described in detail by Gebhardt \etal\ (2021, in prep).   To summarize: HETDEX reductions involve three types of calibration frames:  biases (taken nightly), pixel flats (taken yearly using a laser-driven light source), and twilight sky flats (taken nightly and averaged monthly). The basic steps in the reduction --- bias subtraction, bad pixel masking, fiber profile tracing, wavelength calibration, scattered light removal, spectral extraction, fiber normalization, spectral masking, and sky subtraction --- use these frames, along with the sky background on the science images, to produce a wavelength calibrated, sky-subtracted spectrum for each fiber in the array.

Astrometric calibrations are achieved by measuring the centroid of each field star from fiber counts between 4400~\AA\ and 5200~\AA\ and comparing their positions on the IFUs to the stars' equatorial coordinates in the Sloan Digital Sky Survey \citep[SDSS;][]{York_2000, Abazajian_2009} and \textit{Gaia} \citep{Gaia_2018} catalogs.  This process typically results in global solutions which are good to $\sim 0\farcs 2$, with the exact precision of a measurement dependent upon the number of IFUs in operation at the time of the observation.  

The absolute fluxes for the HETDEX \lya detections are not determined through comparisons with spectrophotometric standard stars. Instead, each HETDEX observation has its own unique flux calibration based on the color and magnitudes of the $g < 24$ SDSS stars in the field. Specifically, HETDEX uses each star's $ugriz$ magnitudes \citep{Padmanabhan_2008}, \textit{Gaia} parallax-based distance \citep{Gaia_2018}, and (very small) foreground reddening \citep{Schlegel_1998, Schlafly_2011} to select its most likely spectral energy distribution from an absolute magnitude-surface gravity-metallicity grid of model stellar spectra \citep{Cenarro_2007, Falcon-Barroso_2011}.  By comparing the most likely flux distributions of $\gtrsim 20$ field stars to their VIRUS spectra, the response curve for any frame can be obtained to a precision of $\sim 5\%.$  Since this accuracy depends on the number of $g < 24$ stars in the field, observations taken with a larger number of operating IFUs achieve greater photometric precision.  Complete details about all these procedures can be found in Gebhardt \etal\ 2021 (in prep). 

The HETDEX spectra were extracted at the \threedhst\ position of each ELG using an optimal weighting algorithm \citep{Horne_1986} that employs a \citet{Moffat_1969} spatial profile set to the seeing conditions of the observation, with a masking aperture set to 3\arcsec.  Each extracted spectrum includes the calibrated flux (in units of $10^{-17}$ ergs~cm$^{-2}$~s$^{-1}$~\AA$^{-1}$), an error array, and a fiber coverage array, which contains the fraction of light each HETDEX fiber collects from a given point source.  Sources whose fiber coverage arrays only contain pixels indicating less than 5\% coverage are excluded from the analysis.  Simulations demonstrate that such a procedure produces an unbiased estimate of the galaxies' emission-line fluxes, though the small astrometric uncertainties associated with the centroids of our extraction apertures make individual line flux measurements only accurate to $\sim 20\%$ (see Appendix~\ref{sec:appendix}).  More information on the extraction method is included in Zeimann \etal\ (2021, in prep).

\begin{figure}[h]
    \centering
    \includegraphics[width=0.45\textwidth]{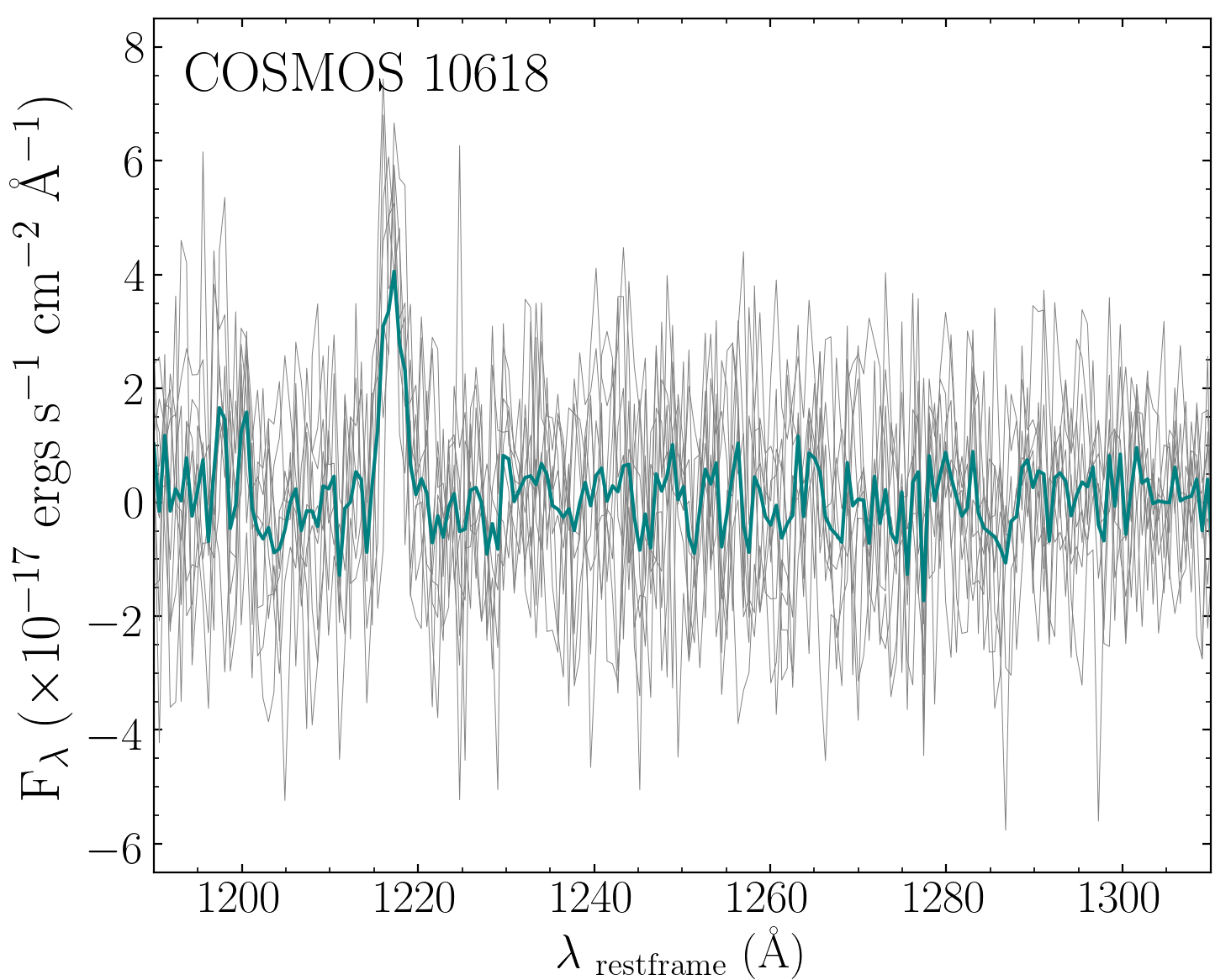}
    \caption{The weighted average spectrum of COSMOS object 10618 formed from all of the available HETDEX data. The nine contributing individual spectra are depicted in light gray, while the weighted average spectrum is shown in teal.  \lya is easily detected.}
    \label{fig:stack}
\end{figure}

Finally, for galaxies with multiple HETDEX observations, a mean spectrum was calculated by weighting each data point by its inverse variance. All spectra were then shifted into the rest frame based on its 3D-HST redshift.  An example of a combined spectrum for an object with nine HETDEX observations is shown in Figure~\ref{fig:stack}.

\section{Computing the \lya Escape Fraction}
\label{sec:fesc}
The physics of the hydrogen atom is well understood.  Under Case B recombination, every ionization produces a Balmer line photon, with $\sim 11.5\%$ of these photons produced by decays from the $n = 4$ state, i.e., \Hb \citep{Pengelly_1964, Osterbrock_2006}. Once an electron reaches the $n=2$ level, it can either remain trapped in the singlet 2S orbital or, if it enters the triplet 2P state, decay via the emission of \lya. In the absence of collisional redistribution, \lya should therefore be produced $\sim 75\%$ of the time.  The intrinsic \lya to \Hb ratio would then be
\begin{equation}
    \frac{I(\mathrm{Ly\alpha})}{I(\mathrm{H\beta})} \sim \frac{3}{4} \frac{\alpha_B}{\alpha_{\rm H\beta}^{\rm eff}} \frac{h\nu_{\mathrm{Ly\alpha}}}{h\nu_{\rm{H\beta}}} \sim 25
\end{equation}
where $\alpha_B$ is the Case B recombination coefficient and $\alpha_{\rm H\beta}^{\rm eff}$ is the effective recombination coefficient for H$\beta$.  In reality, more detailed calculations of the cascade matrix show that under the low-density conditions of the interstellar medium, the production of Ly$\alpha$ is slightly less than this, with $\sim 68\%$ of $n=2$ electrons falling into the 2P state \citep{Hummer_1987, Dijkstra_2014}.  Since the ratio of $\alpha_B$ to $\alpha_{\rm H\beta}^{\rm eff}$  (8.54) is virtually independent of the ISM's electron temperature, under normal circumstances a 100\% escape fraction would mean that \lya is 23 times the brightness of H$\beta$. We therefore define the apparent \lya escape fraction as
\begin{equation}
   f_{\rm esc}^{\rm Ly\alpha} \approx \frac{I(\rm Ly\alpha / \rm H\beta)_{\rm obs}}{23}
   \label{eq:fesc calc}
\end{equation}

Under some conditions, the ratio of 23 is a lower limit: for example, if the interstellar medium becomes optically thin to Lyman continuum photons, the \lya/\Hb ratio may be boosted to values as large as $\sim 33$.  Such an increase is unlikely, as most galaxies at that epoch (even those with strong \OIII\  emission) allow no more than a few percent of Lyman continuum photons to escape into intergalactic space \citep[e.g.,][]{Rutkowski_2017, Naidu_2018, Fletcher_2019}.  Alternatively, if the electron density in star-forming regions becomes sufficiently large ($n_e / \sqrt{T_e/10^4} \gtrsim 10^4$~cm$^{-3}$) collisions will begin to redistribute some fraction of the metastable 2S electrons into the 2P state, again enhancing \lya relative to \Hb. There is little evidence for this effect, as in both the local and $z \sim 2$ universe, most star-forming regions have densities well below this limit \citep[e.g.,][]{Shirazi_2014, Shimakawa_2015, Sanders_2016}.  Finally, strong shocks can increase \lya relative to \Hb by creating an environment where the $n=2$ state of neutral material is collisionally populated.  As a result, in extreme cases, values of \fesc based on a comparison to Balmer emission may be overestimated.  Despite these caveats, for the vast majority of our $z \sim 2$ star-forming galaxies, the assumption that the intrinsic strength of \lya is 23 times that of \Hb should be valid.

\subsection{Corrections to \Hb and \lya}
\label{sec:hb dust}
According to equation (\ref{eq:fesc calc}), the ratio of \lya to \Hb yields \fesc, the apparent escape fraction of \lya photons.  But as written, this value is only an upper limit, as it does not consider \Hb photons that are produced in the galaxy but not detected on earth.  Two physical processes can change what we derive for \Hb.

The first is underlying \Hb absorption in the \threedhst\ grism spectra.  In a population where A-stars are important contributors to the SED, strong Balmer line absorption can eat into a galaxy's \Hb flux and significantly reduce its observed strength.  In the local universe, this effect typically decreases \Hb equivalent widths by $\sim 4$~\AA\ and leads to incorrect estimates of Balmer line decrements \citep{Groves_2012}.  However, the emission-line selected galaxies in our sample have star formation rates and specific star formation rates that are much higher than those found locally \citep{Bowman_2019}, implying the actual \Hb corrections are smaller than this.  Moreover, even if one adopts the local number, \citet{Zeimann_2014} have shown that for $1.90 < z < 2.35$ \threedhst\ emission-line galaxies, the effect of Balmer absorption would only reduce the measured strength of \Hb by $\sim 10\%$. Finally,
we measure \Hb fluxes from the spectra after subtracting the continuum models computed by the \threedhst\ team.  These models were derived from the galaxies' SED fits and thus, to first order, already take the effect of \Hb absorption into account.  
We therefore apply no additional correction for the effect.

A more important process to consider is the attenuation of \Hb by interstellar dust.  While dust is the principle reason for \fesc $< 1$, its presence can also cause \Hb to be underestimated and \fesc to be over-estimated.  If the \threedhst\ spectra extended to H$\alpha$, corrections for the effect of attenuation would be straightforward, since under Case B recombination, the intrinsic H$\alpha$/H$\beta$ line ratio is $\sim 2.86$ \citep{Pengelly_1964, Hummer_1987}.  However, since H$\alpha$ is not available, the extinction of \Hb must be inferred from other methods.

One approach is to use the attenuation measured for the galaxies' stars.  Although the correlation between stellar and nebular attenuation has significant galaxy-to-galaxy scatter, the mean relationship between the two measurements has been quantified many times in both the near and distant universe (e.g.,  \citealp{Calzetti_2000, Battisti_2016, Kashino_2013, Price_2014}; see the table in \citealp{Shivaei_2020} for a complete list).  Thus, in the mean, we can use the galaxies' SEDs to obtain a correction to the observed \Hb fluxes.

To estimate the \Hb extinction, we adopt the stellar reddening estimates derived from rest-frame UV through IR SED-fitting (\citealt{Bowman_2019}; \citeyear{Bowman_2020})  These values are based on age-binned star-formation rate histories similar to those suggested by \citet{Leja_2019}, the \citet{Noll2009} generalization of the \citet{Calzetti_2000} reddening law, and the assumption that the reddening that affects recent star forming regions ($t < 10^7$~yr) is greater than that which attenuates older stellar populations.

Following the results of \citet{Reddy_2020}, who analyzed the Balmer decrements and SEDs of a sample of $\sim 500$ star-forming galaxies between $1.4 < z < 2.6$, we therefore assume 
\begin{equation}
    E(B-V)_{\rm nebular} = 2.07 \times E(B-V)_{\rm stellar}
    \label{eq:nebular-continuum}
\end{equation}
and
\begin{equation}
    A_{{\rm H}\beta} = 3.60 \times E(B-V)_{\rm nebular}
    \label{eq:dust correction}
\end{equation}

Finally, there is one process that may cause us to underestimate the escape of Ly$\alpha$: the scattering of photons in the circumgalactic medium.  Deep imaging and IFU spectroscopy have shown that galaxies are often surrounded by Ly$\alpha$ halos that extend far past the observed extent of their UV continuum \citep[e.g.,][]{Steidel_2011, Momose_2016, Leclercq_2017}.  Since our HETDEX Ly$\alpha$ spectroscopy uses a fixed aperture, it is possible that our \lya flux measurements are missing part of this diffuse, low surface brightness emission.  

Fortunately, the effect of extended \lya halos on our HETDEX spectrophotometry is likely to be minor.  At our sample's median redshift of $z_m = 2.12$, the $3\arcsec$ extraction radius used on the HETDEX spectra corresponds to 24.9~kpc.  This is twice the e-folding scale length of the typical $z = 3.1$ Ly$\alpha$ halo seen by \citet{Matsuda_2012}. Moreover, data from the Multi-Unit Spectroscopic Explorer (MUSE) on the VLT demonstrate that, while small ($1\arcsec$) aperture measurements may miss half the Ly$\alpha$ flux from a typical $z \gtrsim 3$ galaxy, that system's   Ly$\alpha$ curve-of-growth will generally asymptote out by the time it reaches the limits of our $3\arcsec$ aperture \citep{Wisotzki_2016}.  Since the image scale at $z \sim 2$ is only 8\% larger than at $z \sim 3$ and since  Ly$\alpha$ halos are likely smaller at lower redshift \citep{Hayes_2013, Guaita_2015, Wisotzki_2016}, this suggests that our HETDEX apertures encompass most, if not all of the Ly$\alpha$ flux from the bulk of the $z \sim 2$ galaxy population.

\begin{figure*}
  \begin{centering}
    \subfloat[]{
    \includegraphics[height=10.15cm]{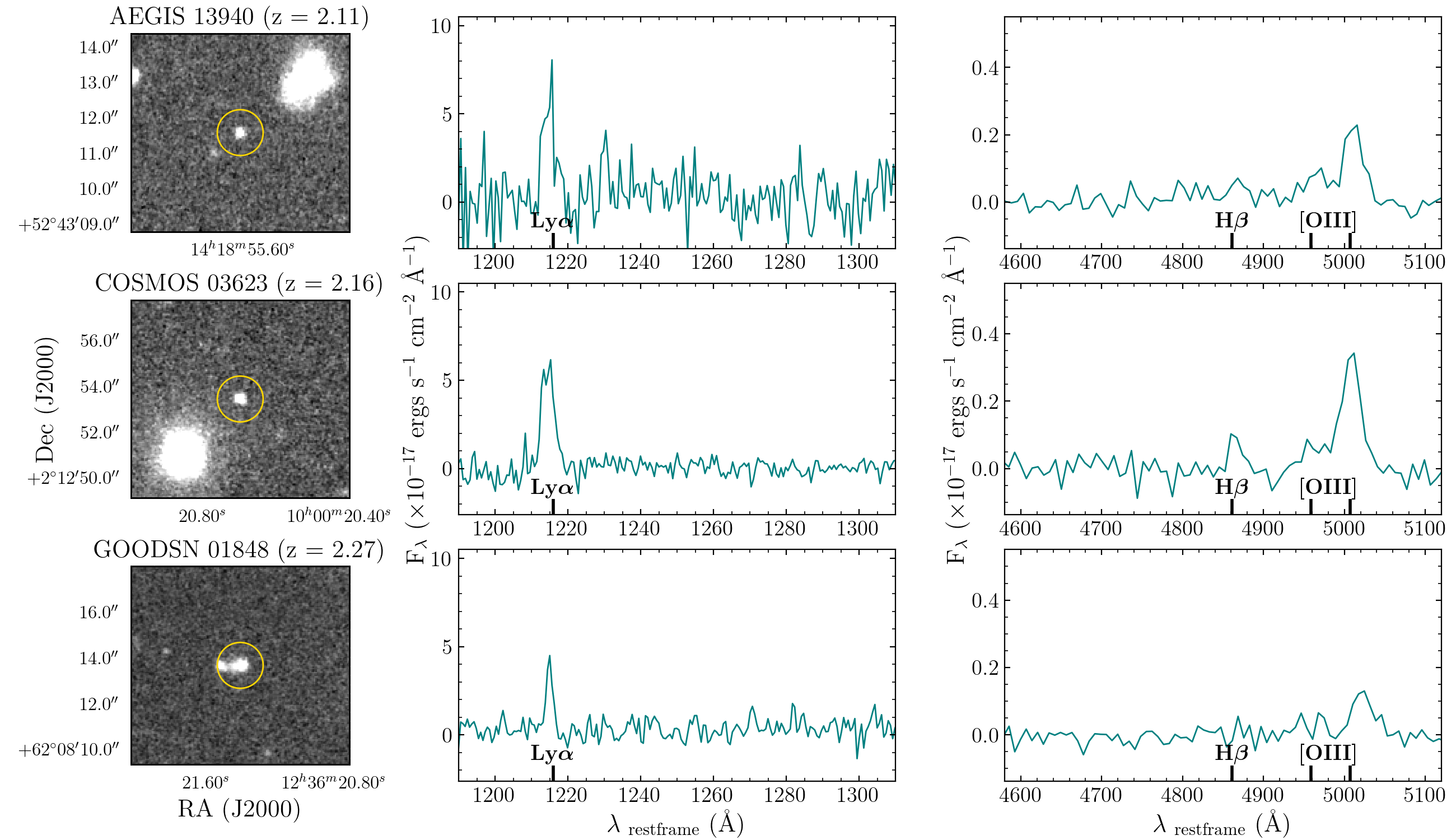}}
  \end{centering}
  \caption{Images and spectra of three \threedhst\ galaxies with strong \lya emission.  The left-hand column shows the galaxy (centered within a $1\arcsec$ circle) as it appears through the WFC3~F160W filter.  The center column displays a segment of the galaxy's HETDEX spectrum: the AEGIS~13950 and GOODS-N~01848 plots show data from a single HETDEX observation, while the spectrum of COSMOS~03623 is the combination of 15 separate pointings.  The right-hand column gives the WFC3~G141 grism spectrum about \OIII\ and \Hb.  These are not typical objects:  in the vast majority of $z \sim 2$ ELGs, \lya is either not present or barely detectable.} \label{fig:examples}
\end{figure*}

\section{Results}
\label{sec:results}

\subsection{\lya from Individual Galaxies}
\label{subsec:singleobjects}
Figure~\ref{fig:examples} presents the spectra of three of our \threedhst\ galaxies.  The figure contains two important features of note.  The first is the weakness of the \Hb line: only $\sim 22\%$ of the \citet{Bowman_2019} sample of \threedhst\ galaxies have \Hb detected with a signal-to-noise ratio above 3, and only 5\% are detected with a signal-to-noise ratio above 5.  This severely constrains our ability to predict the intrinsic strength and the escape fraction of \lya.

The second property displayed in Figure~\ref{fig:examples} is the strength of \lya.  The galaxies shown in the figure are \textit{not} typical of our sample. In most of our objects, \lya is weak or undetected:  in fact,
only 37 of the \nsample\ \threedhst\ objects with HETDEX spectroscopy are present in version 2.1 of the HETDEX emission line catalog (monochromatic
flux limit of $\sim 8 \times 10^{-17}$~ergs~cm$^{-2}$~s$^{-1}$ for \lya at $z \sim 2$).  As a result, any analysis of the escape of \lya from individual galaxies would necessarily be dominated by non-detections.

Because of the large uncertainties associated with the \threedhst\ \Hb measurements and the lack of \lya detections in the HETDEX spectra, a comprehensive study of the escape of \lya from individual galaxies is impossible.  We can, however, perform a stacking analysis of the data.  By binning the galaxies by their physical properties, we can investigate the systematics of \lya escape as a function of galaxy mass, SFR, size, and various other physical parameters. 



\subsection{Stacking Methodology}
\label{subsec:stacking}

The stacking of \lya emission from different galaxies can be challenging.  Due to the resonant scattering of \lya photons through the interstellar and circumgalactic medium, the redshift of a galaxy's \lya emission may be offset from the object's systematic velocity by a considerable amount.  For UV-bright (Lyman break) galaxies at $2 \lesssim z \lesssim 3$, this offset is of the order of $\sim 300$~km~s$^{-1}$ and the \lya line width can be quite broad, $\sim 450$~km~s$^{-1}$ \citep[e.g.,][]{Shapley_2003, Berry_2012}; for fainter systems (i.e., \lya emitters), the velocity difference is smaller ($\sim 200$~km~s$^{-1}$), as is the line's full width at half maximum \citep[e.g.,][]{Hashimoto_2013,Shibuya_2014, Song_2014, Trainor_2015, Muzahid_2020}.   As a result, the co-added \lya line produced from a stack of massive, high SFR galaxies may be significantly wider than that of a similar stack from fainter, lower-mass systems.  This behavior could make the emission line more difficult to measure and introduce a systematic error into our analysis.

One approach to mitigate the systematics associated with line width is to smooth the \lya spectra with a kernel that not only accounts for the behavior of \lya, but also compensates for the limited velocity resolution of the WFC3's G141 grism.  Our \threedhst\ redshifts are somewhat uncertain: based on the 67 emission-line galaxies with both \threedhst\ and ground-based spectroscopy,  the normalized median absolute deviation (NMAD) between the two redshift determinations is $\sigma_{\rm NMAD} = 0.002 (1 + z)$.  This dispersion means that the location of \lya in the 2\,\AA~pixel$^{-1}$ VIRUS spectra is only known to $\pm 3.5$ pixels, or $\sim 600$~km~s$^{-1}$ in the rest-frame of \lya.  This number, when added in quadrature to a possible $\sim 300$~km~s$^{-1}$ kinematic shift in wavelength, suggests that smoothing with a $\sim 670$~km~s$^{-1}$ kernel should minimize the effect that systematic changes in the \lya line widths have on our measurements.  Moreover, while the use of this smoothing kernel does lower the signal-to-noise ratio of our \lya detections, our galaxy sample should still be large enough to enable a robust measurement of the line in the co-added spectra.  

\begin{figure*}
  \centering
  \subfloat[]{%
    \includegraphics[height=8.7cm]{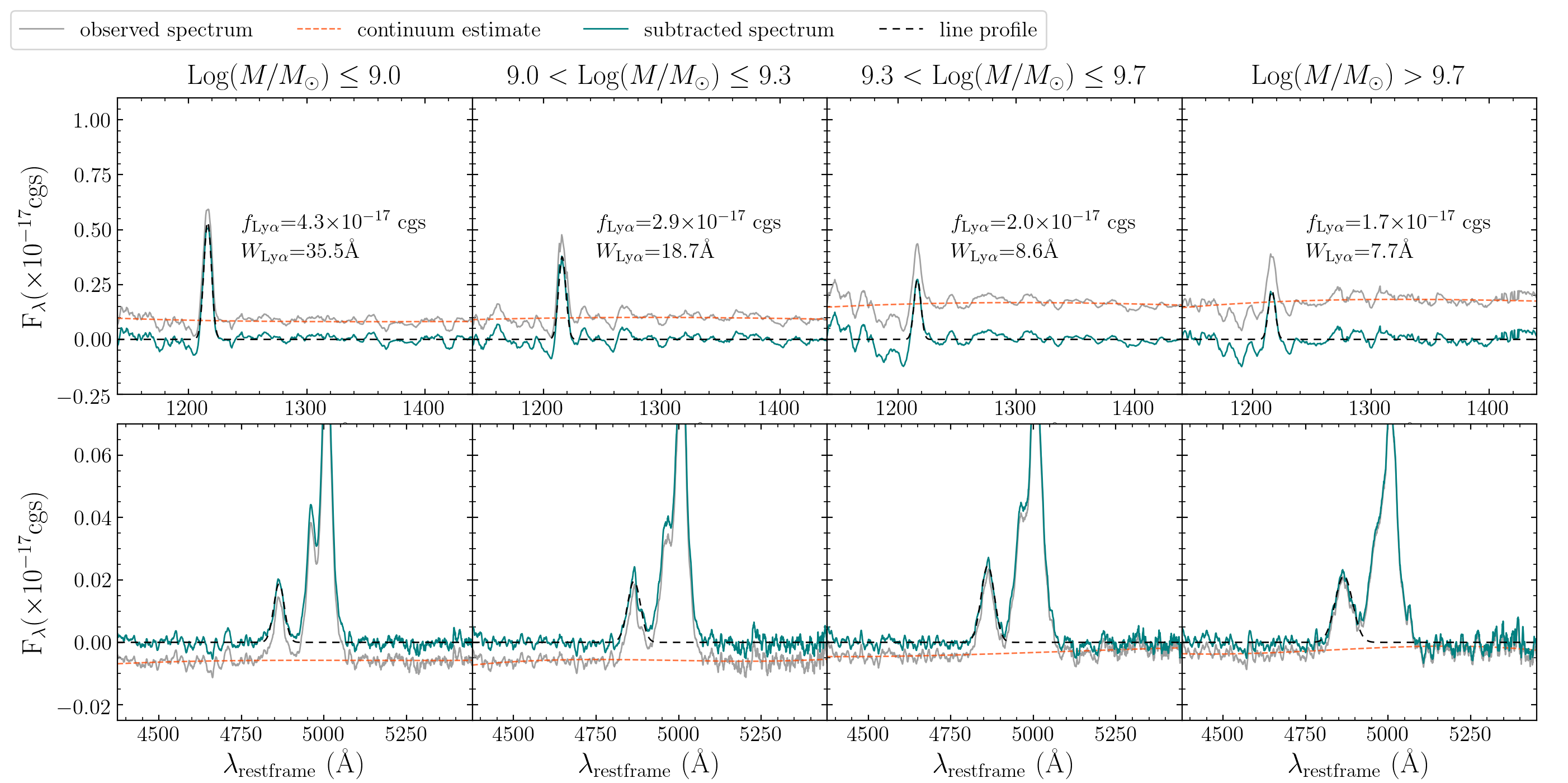}
  }
  \caption{The top panels show stacked \hetdex\ spectra in four stellar mass bins covering the region around \lya; the bottom panels present the corresponding stacks of the spectral region near \Hb and the \OIII\ doublet, as recorded by \threedhst.  In each panel, the grey line represents the original stacked spectrum, the red line is the continuum model, the blue/green line is the continuum-subtracted spectrum, and the dashed black line is the fitted Gaussian model.  As expected, the continua of the stacked \hetdex\ spectra increase with stellar mass; the \lya absorption blueward of the emission line is therefore easier to see in the higher-mass stacks.  The continuum fits in the stacked \threedhst\ spectra are always less than zero, due to the fact that the individual spectra are continuum-subtracted prior to stacking using models that overestimate the continua of objects with strong emission lines \citep{Momcheva_2016}.  This effect is also present in the SED fits of \citet{Bowman_2020}. }
  \label{fig:fits}
\end{figure*}

To create the stacks, we sub-divided our sample evenly into several bins based on galaxy properties, such as stellar mass or star formation rate.  The bin sizes were driven by two requirements: our desire to have a roughly equal number of objects within each bin, and the necessity of reducing the systematic errors associated with the co-addition.  We shifted the spectrum of each galaxy into the rest frame using its \threedhst\ redshift, smoothed the HETDEX data with a $\sigma = 670$~km~s$^{-1}$ Gaussian kernel, and stacked the galaxies within each bin using Tukey's robust biweight estimator \citep{Andrews_1972, Beers_1990}. 
To measure the \lya flux in our stacked spectra, we masked out the emission line, fit a spline to the surrounding region, and used the spline to subtract off the spectral continuum.  We then fit the \lya emission line with a Gaussian and integrated the Gaussian to determine the total line flux.  Our simulations show that this procedure produces an unbiased estimate of the average \lya flux which is good to $\sim 0.3 \times 10^{-17}$~ergs~cm$^{-2}$~s$^{-1}$ (see Appendix~\ref{sec:appendix}).


The \Hb fluxes were measured in a similar manner.  Prior to stacking, we removed the continuum of each galaxy by subtracting off the continuum model computed by \threedhst\ via SED fitting.
However, this procedure was imperfect, as the EAZY SEDs from \cite{Brammer_2008} generally underpredict the strengths of high-excitation forbidden lines such as \OIII\ $\lambda 5007$ \citep{Momcheva_2016, Bowman_2020}.  The mismatch resulted in the galaxies' continua being slightly overestimated, which, in turn, caused our stacks to have a negative continuum level.

To compensate for this issue, we masked out the \Hb and the \OIII\ doublet, fit the \threedhst\ spectra between 4400 and 5500~\AA\ with a spline, and subtracted the residual stacked continuum underlying the emission lines.  We then fit \Hb, \OIII\ $\lambda 4959$, and \OIII\ $\lambda 5007$ with three Gaussians \citep[with the ratio of $\lambda 5007$ to $\lambda 4959$ fixed at 2.98;][]{Storey_2000} and computed the total flux in each line via the integral of these fits. Example emission line fits for the HETDEX and \threedhst\ spectra are shown in Figure~\ref{fig:fits}.  Finally, we used the \lya and \Hb fluxes to calculate the value of \fesc within each bin.


Since we are not able to reliably measure \lya and \Hb in individual galaxies, our approach for estimating the \lya escape fraction differs slightly from those derived by other authors.  Specifically, our escape fractions are determined using the ratio of the biweight \lya to the biweight \Hb, with both lines measured from stacked spectra.  Most other measurements of the $z \gtrsim 2$ \lya escape fraction either come from deriving a mean from a set of individual \fesc measurements \citep[e.g.,][]{Blanc_2011, Oyarzun_2017} or integrating the \lya and \Hb luminosity functions \citep[e.g.,][]{Hayes_2010, Ciardullo_2014}.  This difference should not affect the trends seen in the data.

As detailed in Section~\ref{sec:hb dust}, the escape fractions computed above represent upper limits, as we are likely underestimating the true strength of \Hb due to internal extinction in the galaxies.  To correct for this issue, we de-reddened the individual spectra contributing to the \threedhst\ stacks according to \cite{Reddy_2020}, using each galaxy's $E(B-V)_{\rm nebular}$, as estimated in Section~\ref{sec:hb dust}; this correction increases almost linearly with log stellar mass (see Figure~\ref{fig:fesc_mass_f}).  We then stacked the de-reddened \threedhst\ spectra as before, re-measured \Hb, and calculated (\fesc)*, the dust corrected value of the \lya escape fraction. To obtain the uncertainties on both the raw and dust-corrected escape fractions, we performed a bootstrap analysis on the data, re-sampling (with replacement) the galaxies within each bin, re-stacking the spectra, and then measuring the realization's \lya and \Hb fluxes. After performing 500 such simulations, we defined the 16th and 84th percentiles of the distribution as the $1\sigma$ confidence interval of our measurement.  

\subsection{Results from Stacking}\label{subsec:stacking_results}
To explore the systematics of \lya escape, we began by evenly dividing our dataset into four bins of log stellar mass. The stacked HETDEX and 3D-HST spectra for these bins are shown in Figure~\ref{fig:fits}.  From the figure, it is immediately obvious that high-mass galaxies have smaller \lya escape fractions than their lower mass counterparts, with the strength of \lya declining by more than a factor of $\sim 2$ over the $\sim 1$~dex range in stellar mass.  Also seen in the figure is the presence of \lya absorption on the blue side of the emission line.  This absorption is more noticeable in the higher mass bins where the galactic continua are generally brighter, but a characteristic depression is seen in all the stacks.  This ubiquity suggests that galactic winds are important throughout our sample of star-forming $z \sim 2$ galaxies.  Unfortunately, the feature is too weak and our spectra have too low a spectral resolution for further deconstruction.  In what follows, we simply assume that this absorption is associated with the galaxy as a whole and its effect contributes to our measurement of \fesc. 

For reference, we include \lya flux and rest-frame \lya equivalent width ($W_{\rm Ly\alpha}$) measurements in Figure~\ref{fig:fits}. To measure $W_{\rm Ly\alpha}$, we used \mcsed\
to estimate the continuum component of the best fit SED models, which we then stacked within each mass bin as described in \ref{subsec:stacking}. Following the procedure of \cite{Kornei_2010}, we measured the continuum stack slightly redward of \lya and divided our \lya fluxes by these values to estimate $W_{\rm Ly\alpha}$. These rest-frame equivalent widths range from $\sim 8$ to 36\,\AA, with the equivalent width decreasing with increasing stellar mass, consistent with the results of \cite{Xinnan_2018}. For reference, at $z \sim 3$,  \cite{Kornei_2010} measured the median equivalent width of Lyman-break galaxies with $9 \lesssim \log M/M_{\odot} \lesssim 11$ to be $\sim$ 4~\AA.

\begin{figure}[h]
    \centering
    \includegraphics[width=0.45\textwidth]{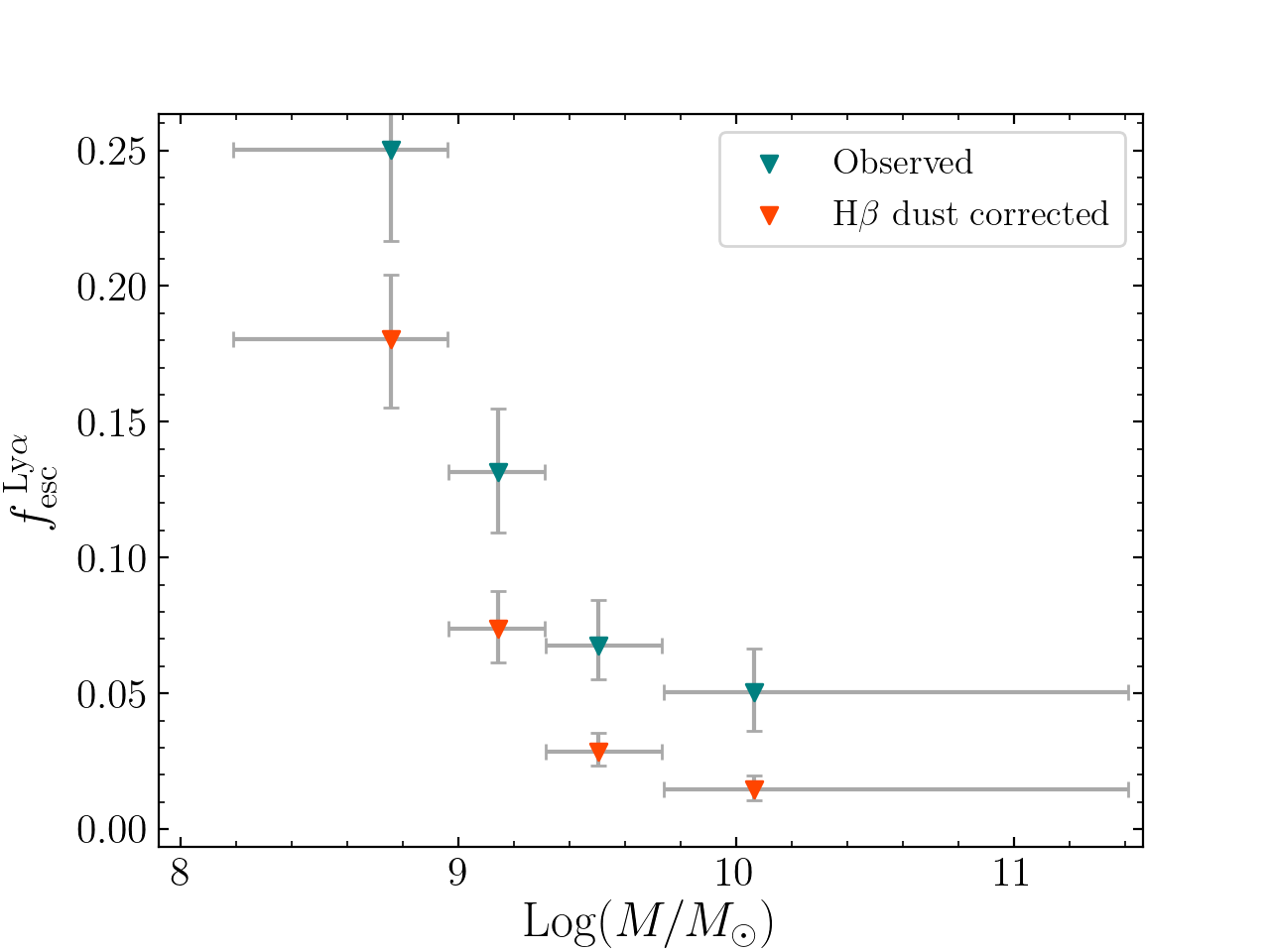}
    \caption{The \lya escape fraction versus log stellar mass. The apparent escape fractions, \fesc, are indicated in teal; our extinction corrected measurements (\fesc)* are shown in orange. The error bars in the escape fraction are determined from bootstrapping; those in mass represent the sizes of the bins.  The anti-correlation is the same as that seen by \citet{Oyarzun_2017}.}
    \label{fig:fesc_mass_f}
\end{figure}

Our values for \fesc and (\fesc)* in the four mass bins are plotted in Figure~\ref{fig:fesc_mass_f} and summarized in Table~\ref{tab:fesc_mass_t}.  Table~\ref{tab:fesc_mass_t} also lists the number of spectra contributing to each stack ($N_{\rm stack}$) and the median value of $c_{\rm H\beta} = A_{\rm H\beta} / 2.5$ for the stacked galaxies, using $A_{\rm H\beta}$ values computed via equation (\ref{eq:dust correction}).  Overall, the escape fraction of \lya declines steadily from $\sim 18\%$ in galaxies with $\log M/M_{\odot} \lesssim 9$ to $\sim 1\%$ for systems with $\log M/M_{\odot} \gtrsim 9.7$.  This result is similar to that reported by \citet{Oyarzun_2016} and \citet{Oyarzun_2017}, who derived \fesc for $3 < z < 4.6$ \threedhst\ galaxies using the sources' star formation rates.

\begin{deluxetable}{lc lc lc lc lc}[h]
\tablecaption{\fesc vs Stellar Mass \label{tab:fesc_mass_t}}
\tablewidth{0pt}
\tablehead{
\colhead{Mass Bin} & \colhead{\fesc} & \colhead{(\fesc)*} & \colhead{$c_{\rm H\beta}$} & \colhead{N$_{\rm stack}$} }
\startdata
 $\log M / M_\odot \leq 9.0$ & $0.25^{+0.03}_{-0.03}$  & $0.18^{+0.02}_{-0.02}$ & 0.12 & 234 \\
 $9.0  < \log M/ M_\odot \leq 9.3$ & $0.13^{+0.02}_{-0.02}$ & $0.08^{+0.01}_{-0.01}$ & 0.23 & 234 \\
 $9.3  <  \log M/ M_\odot \leq 9.7$ & $0.07^{+0.02}_{-0.01}$ & $0.031^{+0.007}_{-0.005}$ & 0.34 & 234 \\
 $\log M/ M_\odot > 9.7$ & $0.05^{+0.02}_{-0.01}$ & $0.016^{+0.005}_{-0.004}$ & 0.47 & 233
\enddata
\end{deluxetable}

\begin{figure*}
    \centering
    \includegraphics[height=5.5cm]{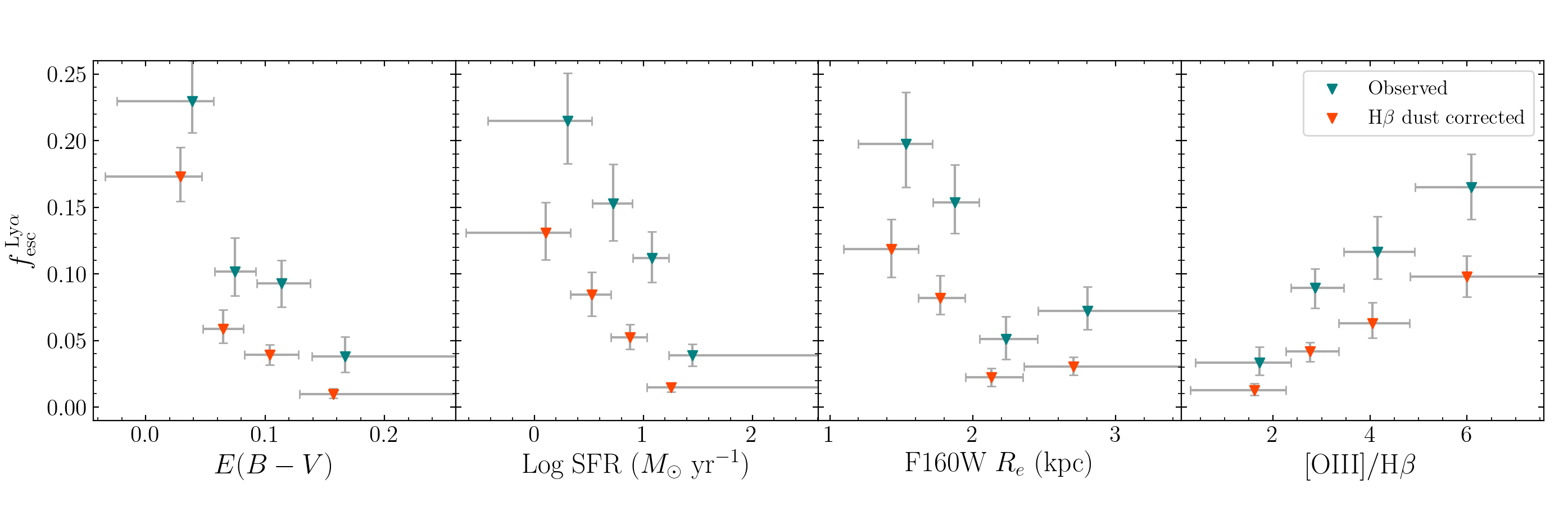}
    \caption{The \lya escape fraction versus $E(B-V)$, log SFR, F160W $R_e$, and \mbox{\OIII\ $\lambda 5007$/\Hb}. The values for \fesc are indicated in teal and the values for (\fesc)* are shown in orange.  The error bars in the escape fraction represent $1\,\sigma$ bootstrapped uncertainties; those in mass illustrate the extent of the bins.  For clarity, the dust-corrected measurements are offset slightly, and the highest value bin's width is truncated. There is a decrease in \fesc and (\fesc)* with increasing $E(B-V)$ and SFR, an increase in \fesc and (\fesc)* with increasing \mbox{\OIII\ $\lambda 5007$/\Hb}, and an overall decrease in \fesc and (\fesc)* with increasing $R_e$.}
    \label{fig:fesc_other_f}
\end{figure*}

Similarly, we divided our sample into bins of $E(B-V)_{\rm stellar}$ (which we will refer to as $E(B-V)$) for simplicity), SFR, rest-frame optical effective radius ($R_e$), and \mbox{\OIII\ $\lambda 5007$/\Hb} ratio, and investigated the behavior that these physical quantities have on the escape of \lya.  The results from these stacks are given in Table~\ref{tab:fesc_other_t} and displayed in Figure~\ref{fig:fesc_other_f}.  Not all of the galaxies have measurements of $R_e$ and \mbox{\OIII\ $\lambda 5007$/\Hb}, due either to issues with the \textit{HST} imaging or low \Hb signal-to-noise ratios.  Still, even our most stringent requirements produce a galaxy sample with more than 650 objects. 

None of the relations shown in Figure~\ref{fig:fesc_other_f} are especially surprising.  The anti-correlation between \fesc and internal extinction is expected, as even a small amount of dust can break the chain of resonant scatterings required for \lya to escape its immediate environment \citep{Ahn_2001, Verhamme_2006, Dijkstra_2006, Rivera-Thorsen_2015}.  Similarly, one might expect \lya to have an easier time escaping small, compact galaxies, as, all things being equal, the photons would undergo fewer scatterings before leaving the galaxy \citep[e.g.,][]{Afonso_2018}. The anti-correlation with star-formation rate is plausible since we expect most $z \sim 2$ emission-line galaxies to lie along a star-forming galaxy ``main sequence''  \citep[e.g.,][]{Noeske_2007, Rodighiero_2011}. Since stellar mass and SFR are correlated, any relationship between \fesc and mass will likely carry over to SFR\null.  Finally, the correlation between \mbox{\OIII\ $\lambda 5007$/\Hb} and \fesc has been seen previously by \cite{Erb_2016}, who attributed the trend to the ratio's dependence on metallicity.  These authors argue that, in addition to having less dust, lower metallicity systems are likely to have hotter stars, which can produce higher ionization parameters in the interstellar medium.  This situation could result in reduced covering fractions or column densities of neutral hydrogen, and facilitate the escape of \lya from its local environment.

\begin{deluxetable}{lc lc lc lc}[h]
\tablecaption{\fesc vs $E(B-V)$, SFR, F160W $R_e$, and \mbox{\OIII\ $\lambda 5007$/\Hb} \label{tab:fesc_other_t}}
\tablewidth{0pt}
\tablehead{
\colhead{Bin} & \colhead{\fesc} & \colhead{(\fesc)*} & \colhead{N$_{\rm stack}$} }
\startdata
 $E(B-V) \leq 0.06$ & $0.23^{+0.03}_{-0.02}$  & $0.18^{+0.02}_{-0.02}$ & 248\\
 0.06 $ < E(B-V)\leq 0.09$ & $0.10^{+0.03}_{-0.02}$ & $0.06^{+0.01}_{-0.01}$ & 222\\
 0.09 $ < E(B-V) \leq 0.14$ & $0.09^{+0.02}_{-0.02}$ & $0.042^{+0.007}_{-0.008}$ & 233\\
 $E(B-V)$ \textgreater~0.14 & $0.04^{+0.01}_{-0.01}$ & $0.011^{+0.004}_{-0.003}$ & 232\\
\hline 
 Bin ($\log M_{\odot}$~yr$^{-1})$ & \fesc & (\fesc)* & N$_{\rm stack}$ \\
 \hline
 SFR $\leq 0.53$ & $0.21^{+0.04}_{-0.03}$ & $0.13^{+0.02}_{-0.02}$ & 234\\
 0.53 $ < $ SFR $\leq 0.89$ & $0.15^{+0.03}_{-0.03}$ & $0.09^{+0.02}_{-0.02}$ & 234\\
 0.89 $ < $ SFR $\leq 1.22$ & $0.11^{+0.02}_{-0.02}$ & $0.06^{+0.01}_{-0.009}$ & 233\\
 SFR \textgreater~1.22 & $0.039^{+0.009}_{-0.008}$ & $0.016^{+0.004}_{-0.003}$ & 234\\
\hline 
Bin (kpc) & \fesc & (\fesc)* & N$_{\rm stack}$ \\
\hline
$R_e \leq 1.72$ & $0.20^{+0.04}_{-0.03}$  & $0.12^{+0.02}_{-0.02}$ & 194\\
1.72 $ < R_e \leq 2.05$ & $0.15^{+0.03}_{-0.02}$ & $0.09^{+0.02}_{-0.01}$ & 193\\
2.05 $ < R_e \leq 2.46$ & $0.05^{+0.02}_{-0.02}$ & $0.023^{+0.007}_{-0.007}$ & 194\\
$R_e$ \textgreater~2.46 & $0.07^{+0.02}_{-0.01}$ & $0.033^{+0.007}_{-0.006}$ & 193\\
\hline
Bin & \fesc & (\fesc)* & N$_{\rm stack}$ \\
\hline
\OIII/\Hb $\leq 2.4$ & $0.03^{+0.01}_{-0.01}$ & $0.014^{+0.005}_{-0.004}$ & 165\\
$2.4 < $ \OIII/\Hb $\leq 3.5$ & $0.09^{+0.01}_{-0.01}$ & $0.044^{+0.007}_{-0.008}$ & 163\\
$3.5 < $ \OIII/\Hb $\leq 4.9$ & $0.12^{+0.03}_{-0.02}$ & $0.07^{+0.02}_{-0.01}$ & 165\\
\OIII/\Hb \textgreater~4.9 & $0.17^{+0.02}_{-0.02}$ & $0.10^{+0.02}_{-0.02}$ & 163
\enddata
\end{deluxetable}

\subsection{What Relations are Fundamental?}
\label{subsec:guang}

Figures~\ref{fig:fesc_mass_f} and \ref{fig:fesc_other_f} demonstrate that the \lya escape fraction varies systematically with stellar mass, dust content, star formation rate, half-light radius, and \mbox{\OIII\ $\lambda 5007$/\Hb} line ratio. However, as the Pearson correlation coefficients of Table~\ref{tab:correlation} demonstrate, many of these variables are interdependent.  Under these conditions, it can be difficult to determine which property is driving a correlation and which are simply confounding variables.  Because the correlations displayed in Figures~\ref{fig:fesc_mass_f} and \ref{fig:fesc_other_f} involve just four points, sophisticated statistical techniques for disentangling the relationships are of limited value. Instead, we can attempt to separate the variables following a procedure similar to one described in \citet{Guang_2017}.


\begin{deluxetable}{lcc}
\tablecaption{Correlation Strengths \label{tab:correlation}}
\tablewidth{0 pt}
\tablehead{
\colhead{Sample Properties} & \colhead{$r$ value}}
\startdata
 $E(B-V)$ vs.\ Log $M/M_{\odot}$  & $+0.76$ \\
 Log SFR vs.\ Log $M/M_{\odot}$   & $+0.62$ \\
 $R_e$ vs.\ Log $M/M_{\odot}$     & $+0.45$ \\
 Log SFR vs.\ $E(B-V)$            & $+0.36$ \\
 \OIII/\Hb vs.\ $E(B-V)$          & $-0.36$ \\
 \OIII/\Hb vs.\ Log $M/M_{\odot}$ & $-0.33$ \\
 Log SFR vs.\ $R_e$               & $+0.32$ \\
 $R_e$ vs.\ $E(B-V)$              & $+0.28$ \\
\enddata
\end{deluxetable}


In short, \citet{Guang_2017} investigated the behavior of black hole accretion as a function of both host galaxy stellar mass and star formation rate. After binning their sample into SFR intervals, they split each bin in two: those with $M >$ med$(M)$ and those with $M <$ med$(M)$, where med$(M)$ is the median stellar mass of the bin.
After measuring the black hole accretion rate within each sub-bin, they flipped test, binning by stellar mass and sub-dividing the bins along the median SFR\null.  The differences between the accretion rates measured from the above and below median sub-samples for a given property demonstrate how closely the property and accretion rate are correlated. This, along with partial correlation (PCOR) analyses, served to determine which host-galaxy property is primarily related to supermassive black hole growth. 
While we cannot replicate their test due to our lack of individual measurements, we can employ a similar logic to explore which properties are driving the trends displayed in Figures~\ref{fig:fesc_mass_f} and \ref{fig:fesc_other_f}.
\begin{figure*}
    \centering
    \includegraphics[height=11cm]{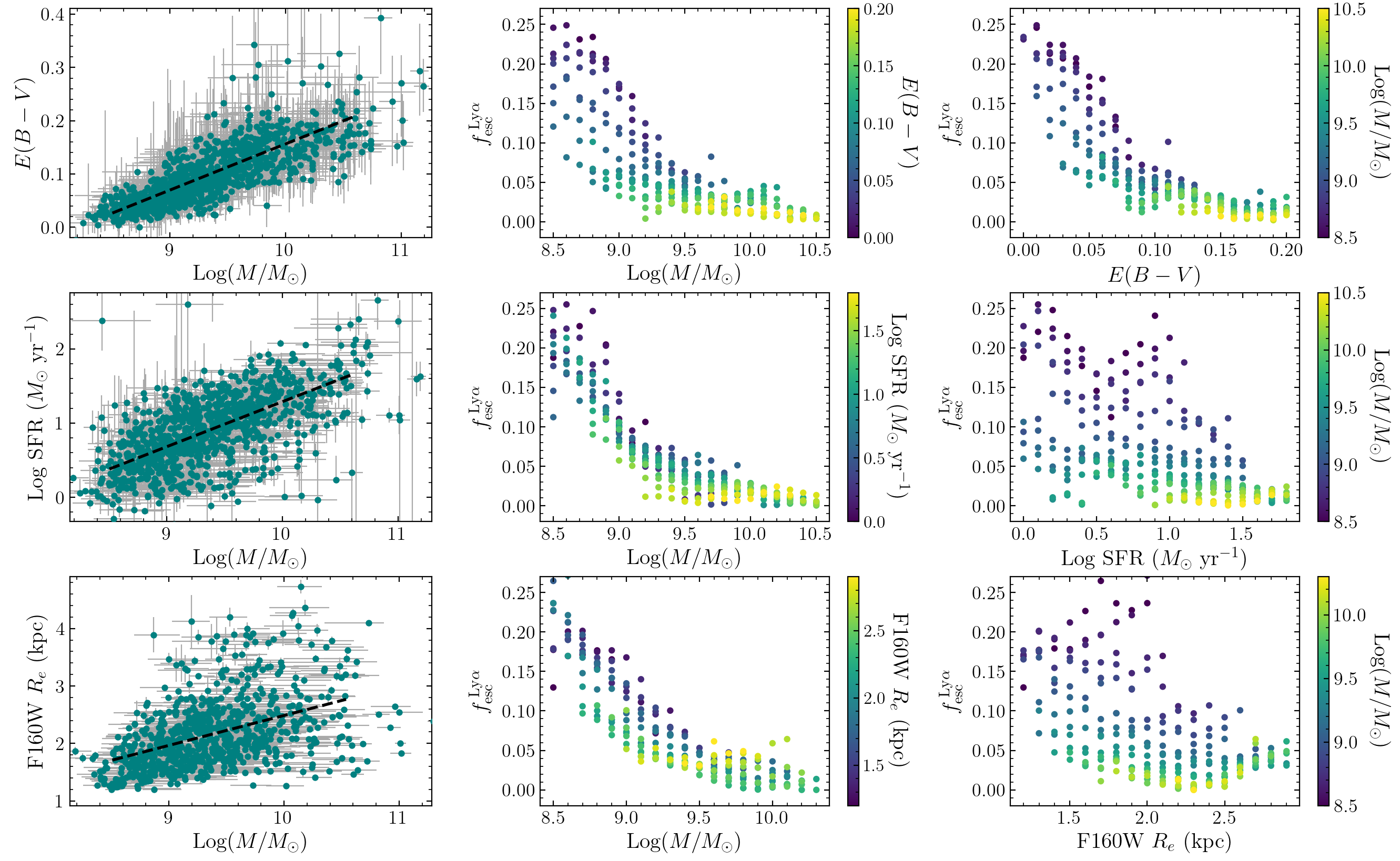}
    \caption{\textit{Left panels, from top to bottom:} The scatter plot of $E(B-V)$, log SFR, and $R_e$ versus log stellar mass for the galaxies being analyzed. The black dashed line indicates a linear least squares regression to the data. \textit{Middle panels, from top to bottom:}  The dust-correction \lya escape fraction \fesc versus log stellar mass as a function of $E(B-V)$, log SFR, and $R_e$. \textit{Right panels, from top to bottom:} The dust-correction \lya escape fraction \fesc versus $E(B-V)$, log SFR, and $R_e$ as a function of log stellar mass. When comparing the middle and right panels of the first row, we see little to no difference in the amount of scatter between the two plots. This suggests that stellar mass and dust attenuation have similar degrees of correlation with \fesc. Comparing the middle and right panels of the second row, we observe greater scatter in the plot of escape fraction versus SFR\null. This shows that, between stellar mass and star formation rate, \fesc better correlates with stellar mass. When comparing the middle and right panels of the third row, we see greater scatter in the plot of \fesc versus $R_e$, implying that between stellar mass and effective radius, the \lya escape fraction is better correlated with stellar mass.}
    \label{fig:greg_tests}
\end{figure*}

We began by comparing the behavior of \fesc with stellar mass and internal extinction, due to the correlation strength of these two variables (see Table \ref{tab:correlation}). We first divided the stellar mass-extinction plane in a $21 \times 21$ grid of values, with the limits of this grid spanning the entire ranges of log($M/M_{\odot}$) and $E(B-V)$ for our sample. For each cell in the grid we defined an elliptical aperture with the semi-major and semi-minor axes fixed to three times the cell width. These values were chosen to encompass a sufficient number of objects while retaining the identity of the cell. We then collected all the objects with stellar masses and extinctions that fell within this aperture, and if this number was greater than 25, we stacked their spectra and measured (\fesc)* using the procedures described in Section~\ref{subsec:stacking}. For simplicity, we only used the \Hb dust-corrected values in this analysis, but we will refer to the measurements simply as \fesc.

The result of this procedure was a 
$21 \times 21$ grid of \fesc values across the log($M/M_{\odot}$)-$E(B-V)$ plane. To investigate whether \fesc is more closely linked to stellar mass or extinction, we plotted \fesc against log($M/M_{\odot}$) as a function of $E(B-V)$, and \fesc against $E(B-V)$ as a function of  log($M/M_{\odot}$). Essentially, we asked the question, ``for galaxies with a given stellar mass, does internal extinction make a difference in predicting \fesc?''  And, conversely, ``for galaxies with a given value of extinction, does stellar mass make a difference in predicting \fesc?'' Smaller scatter indicates a tighter correlation with \fesc.

The results are displayed in the middle and right panels of the first row of Figure~\ref{fig:greg_tests}. Both plots show a strong negative correlation with 
\fesc and similar degree of scatter. In this case, we cannot tell whether 
log($M/M_{\odot}$) or $E(B-V)$ is more closely correlated with the escape of \lya. However, the plots do allow us to more closely examine the trends of \fesc with stellar mass and internal reddening originally seen in Figures~\ref{fig:fesc_mass_f} and \ref{fig:fesc_other_f}. Unsurprisingly, at fixed stellar mass, galaxies with low $E(B-V)$ exhibit the highest values of \fesc, and at fixed $E(B-V)$, galaxies with low stellar mass exhibit the highest values of \fesc. 

We repeated this procedure for the next pair of highly correlated variables: stellar mass and star formation rate. The results are displayed in the middle and right panels of the second row of Figure~\ref{fig:greg_tests}. Both plots show a negative correlation with \fesc, but the plot of \fesc versus log($M/M_{\odot}$) displays significantly less scatter than that for \fesc versus log~SFR\null. This result demonstrates that \fesc is more tightly linked with stellar mass than it is with star formation rate, and the anti-correlation with SFR is likely driven by the existence of the star-forming galaxy main sequence. 

In the middle and right panels of the bottom row of Figure~\ref{fig:greg_tests}, we compare the behavior of
\fesc using two variables that are not as strongly correlated:  stellar mass and galaxy effective radius.  We observe that the plot of \fesc versus log($M/M_{\odot}$) shows less scatter than that for \fesc versus $R_e$, implying that \fesc is more tightly correlated with stellar mass than physical size.  Based on the plots, it appears that size is just a confounding variable.

\begin{figure*}
    \centering
    \includegraphics[height=8.75cm]{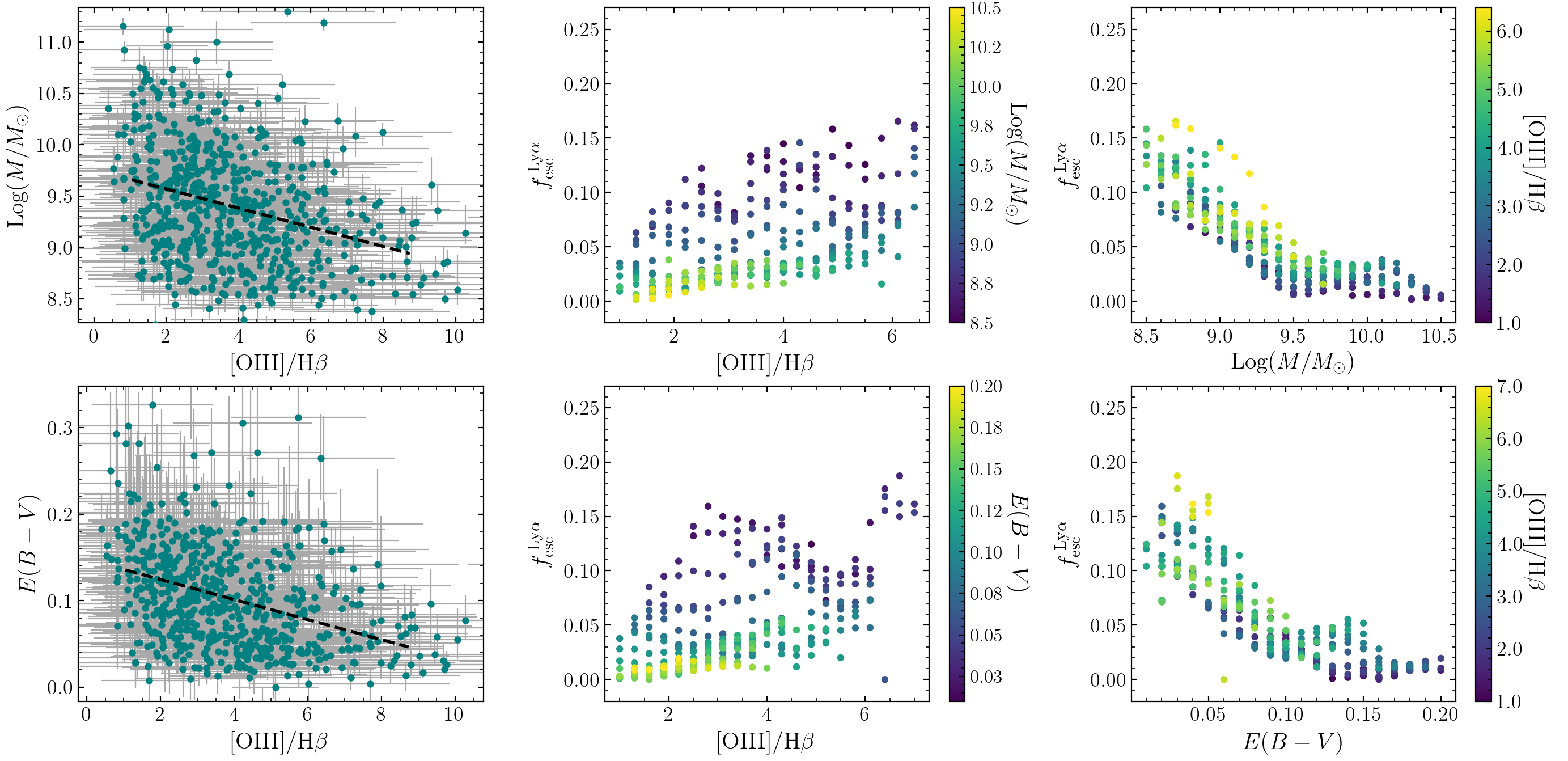}
    \caption{\textit{Left panels, from top to bottom:} The scatter plot of log stellar mass and $E(B-V)$ versus \mbox{\OIII\ $\lambda 5007$/\Hb} for the galaxies being analyzed. The black dashed line indicates a linear least squares regression to the data. \textit{Middle panels, from top to bottom:}  The dust-correction \lya escape fraction (\fesc)* versus \mbox{\OIII\ $\lambda 5007$/\Hb} as a function of log stellar mass and $E(B-V)$. \textit{Right panels, from top to bottom:} The dust-correction \lya escape fraction \fesc versus log stellar mass and $E(B-V)$ as a function of \mbox{\OIII\ $\lambda 5007$/\Hb}. Comparing the middle and right panels of the first row, we observe greater scatter in the plot of escape fraction versus \mbox{\OIII\ $\lambda 5007$/\Hb}. This implies that, between the \mbox{\OIII\ $\lambda 5007$/\Hb} line ratio and stellar mass, \fesc better correlates with stellar mass. When comparing the middle and right panels of the second row, we again see greater scatter in the plot of (\fesc)* versus \mbox{\OIII\ $\lambda 5007$/\Hb}, implying that between the \mbox{\OIII\ $\lambda 5007$/\Hb} line ratio and internal extinction, the \lya escape fraction is better correlated with extinction.}
    \label{fig:greg_tests_oiii_hb}
\end{figure*}

We performed one further \fesc comparison: is stellar mass is more important for predicting \fesc than the \OIII $\lambda 5007$/\Hb ratio?  As first pointed out by \citet{Erb_2016} and confirmed in Figure~\ref{fig:fesc_other_f}, $z\sim 2$ galaxies with extreme line ratios are more likely to emit \lya than the more typical star-forming galaxies of the epoch.  While stellar mass and $\lambda 5007$/\Hb ratio are not highly correlated, we again see that \fesc is more tightly linked with stellar mass. \citet{Erb_2016} have argued that the correlation between extreme line ratios and \fesc is likely due to the line being associated with low stellar metallicity, which generally implies systems with less dust and higher ionization parameters.  While this may be true, the relationship between internal extinction and \fesc is tighter, and since stellar mass is well-correlated with extinction, it is this link that appears to dominate. When we use our analysis to compare the behavior of \fesc with \mbox{\OIII\ $\lambda 5007$/\Hb} and $E(B-V)$, \fesc is indeed more tightly correlated with extinction. The results of these \mbox{\OIII\ $\lambda 5007$/\Hb} comparisons are displayed in the middle and right panels of Figure~\ref{fig:greg_tests_oiii_hb}. 

One caveat to this analysis involves the intrinsic uncertainties of the parameters with respect to the radii defining our elliptical aperture. The scatter in \fesc has two components --- the intrinsic scatter in the galaxy properties and observational error --- and the heteroskedastic nature of the latter makes quantifying its effect on our elliptical apertures difficult to model.  This uncertainty is compounded by small number statistics, as each stack contains significantly fewer objects ($N_{\rm objects} \geq 25$) than in Section~\ref{subsec:stacking_results}. Nevertheless, the trends reveal that \fesc is more closely correlated with stellar mass and $E(B-V)$ than the other properties.

\section{The Integrated Escape Fraction}
\label{sec:global_escape}
Finally, we can use our stacking procedure to estimate the mean \lya escape fraction for our entire data set of 3D-HST emission line galaxies. As discussed in Section~\ref{sec:sample}, at $z \sim 2$, our \threedhst\ sample is primarily selected via the presence of strong \OIII\ emission.  To determine the sample's mean value of (\fesc)* with a minimum of selection effects, we can therefore bin the data by \mbox{\OIII\ $\lambda 5007$}  luminosity, measure each bin's mean \lya escape fraction, and combine the measurements by weighting each bin's escape fraction by the number of galaxies in the bin.

\begin{figure}[h]
    \centering
    \includegraphics[width=0.45\textwidth]{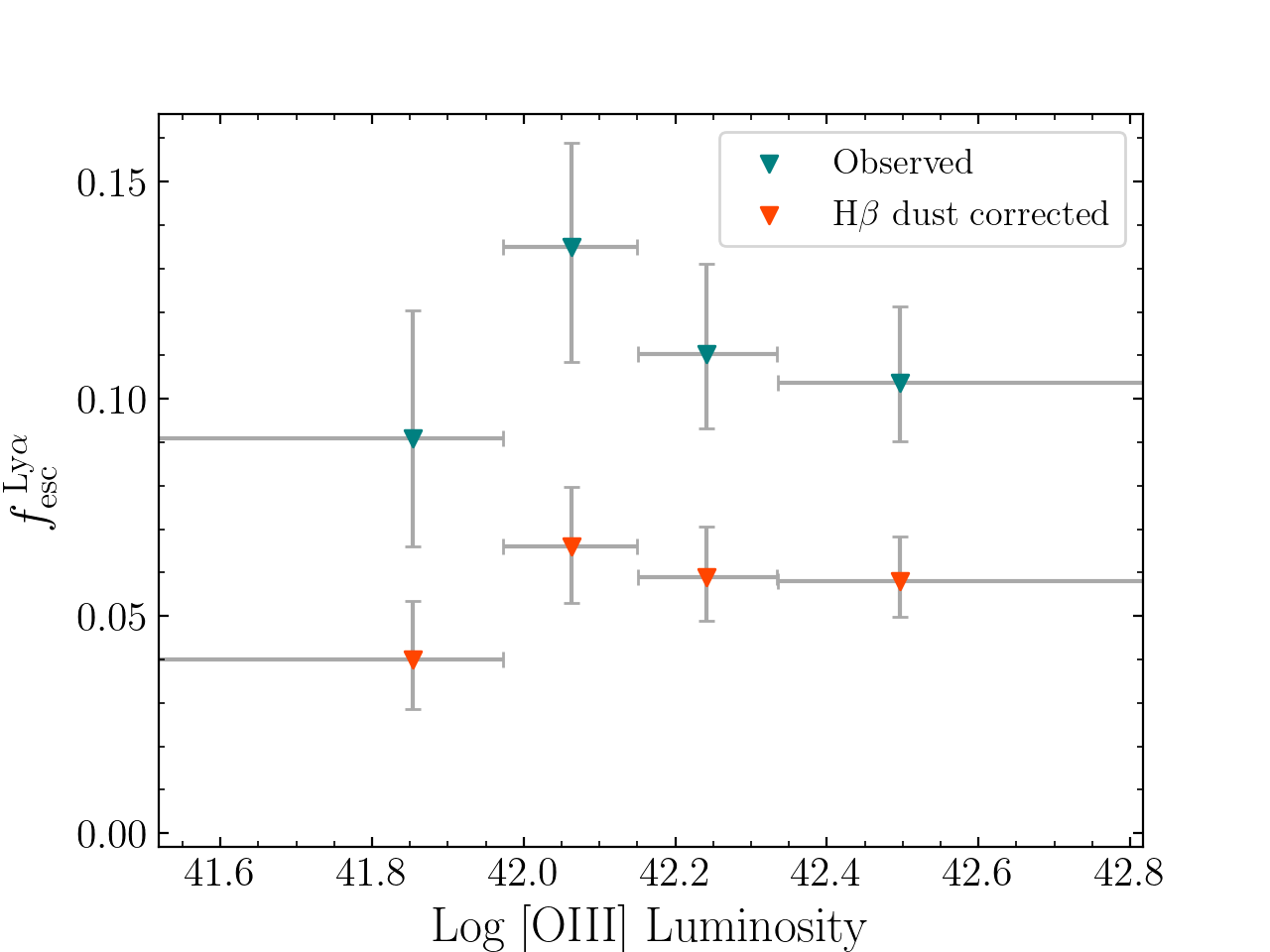}
    \caption{The \lya escape fraction versus log \mbox{\OIII\ $\lambda 5007$} luminosity.  The teal points are taken directly from the measured values of \lya and \Hb and represent upper limits; the orange points are those where \Hb has been corrected for internal reddening.  The error bars in the escape fraction represent  $1\,\sigma$ bootstrapped uncertainties; those in \mbox{\OIII\ $\lambda 5007$} luminosity illustrate the extent of the bins.  Any dependence of the escape fraction on \mbox{\OIII\ $\lambda 5007$} luminosity is weak, and the overall value of (\fesc*) is $\sim 5\%$.}
    \label{fig:OIII_luminosity}
\end{figure}

Figure~\ref{fig:OIII_luminosity} shows our measurements of the \lya escape fraction as a function of \mbox{\OIII\ $\lambda 5007$} luminosity.  Interestingly, the escape fraction is remarkably consistent across the bins, with the weighted average value being (\fesc)* = $5.8^{+0.7}_{-0.5} \%$.   Alternatively, since the numbers are so consistent, we can simply stack the spectra from the entire dataset and estimate the overall fraction of \lya photons escaping from the rest-frame optical emission line galaxies.  This escape fraction of $6.0^{+0.6}_{-0.5} \%$ is consistent with the previous measurements of emission-line galaxies in the $z \sim 2$ universe \citep{Hayes_2010, Ciardullo_2014}.  Of course, our estimate of (\fesc)* does not include the contributions of non-emission lines galaxies, such as systems that are quiescent or heavily extincted.  But the amount of \lya escaping from these galaxies should be small.  Thus, our ELG-based value for the (\fesc)* should be close to the volumetric escape fraction of the epoch.

\section{Conclusion}\label{sec:discuss}
While many hundreds of bright $z\gtrsim 2$ galaxies now have high-quality spectra in the rest-frame UV and optical \citep[e.g.,][]{Strom_2017, Theios_2019, Topping_2020} these objects are generally selected via their colors on deep broadband images.  Here we use a uniform sample of emission-line selected galaxies to investigate the \lya escape fraction.  The difficulties associated with detecting and accurately measuring \lya from individual $z \sim 2$ sources makes the identification of trends in the \lya escape fraction difficult to quantify. However, by co-adding (stacking) the smoothed, low-resolution HETDEX spectra of \nsample\ \OIII\ emitting galaxies between $1.9 < z < 2.35$, we have been able to measure \lya fluxes for galaxy samples where the \lya line is largely weak or undetectable.  These data, when compared to \Hb measurements from stacked \threedhst\ grism spectra, allowed for a determination of the escape fraction of \lya photons as a function of a wide range of galaxy properties using a minimum of external assumptions.  

Our data demonstrate that the fraction of \lya photons escaping a galaxy depends greatly on the type of galaxy being observed.  In general, \fesc and its de-reddened counterpart, (\fesc)*, are inversely correlated with stellar mass, SFR, size, and internal $E(B-V)$, and directly correlated with the strength of \OIII\ $\lambda 5007$ relative to \Hb.  However, not all of these correlations are fundamental.  In particular, the dependence of the \lya escape fraction on SFR is likely indirect, and only observed because SFR correlates with stellar mass. By disentangling these correlations, we determine that galactic stellar mass and dust reddening are the properties with which \fesc are most tightly linked. Future studies aimed at detecting \lya from complete samples of individual galaxies with \Ha/\Hb derived extinction values could yield more insight into which property (if any) is the most fundamental in determining the \lya escape fraction.

\acknowledgments
This work was supported by the NSF through grant AST-1615526 and through NASA Astrophysics Data Analysis grant NNX16AF33G, and was based on observations taken by the CANDELS Multi-Cycle Treasury Program with the NASA/ESA HST, which is operated by the Association of Universities for Research in Astronomy, Inc., under NASA contract NAS5-26555.  The data were obtained from the Hubble Legacy Archive, which is a collaboration between the Space Telescope Science Institute (STScI/NASA), the Space Telescope European Coordinating Facility (STECF/ESA), and the Canadian Astronomy Data Centre (CADC/NRC/CSA).

Observations were obtained with the Hobby-Eberly Telescope (HET), which is a joint project of the University of Texas at Austin, The Pennsylvania State University, Ludwig-Maximilians-Universit\"at M\"unchen, and Georg-August-Universit\"at G\"ottingen. The HET is named in honor of its principal benefactors, William P. Hobby and Robert E. Eberly.

VIRUS is a joint project of the University of Texas at Austin (UTA), Leibniz-Institut f{\" u}r Astrophysik Potsdam (AIP), Texas A\&M University (TAMU), Max-Planck-Institut f{\" u}r Extraterrestriche-Physik (MPE), Ludwig-Maximilians-Universit{\" a}t M{\" u}nchen, The Pennsylvania State University, Institut f{\" u}r Astrophysik G{\" o}ttingen, University of Oxford, Max-Planck-Institut f{\" u}r Astrophysik (MPA), and The University of Tokyo.

HETDEX is led by the University of Texas at Austin McDonald Observatory and Department of Astronomy with participation from the Ludwig-Maximilians-Universit\"at M\"unchen, Max-Planck-Institut f\"ur Extraterrestriche-Physik (MPE), Leibniz-Institut f\"ur Astrophysik Potsdam (AIP), Texas A\&M University, The Pennsylvania State University, Institut f\"ur Astrophysik G\"ottingen, The University of Oxford, Max-Planck-Institut f\"ur Astrophysik (MPA), The University of Tokyo, and Missouri University of Science and Technology. In addition to Institutional support, HETDEX is funded by the National Science Foundation (grant AST-0926815), the State of Texas, the US Air Force (AFRL FA9451-04-2-0355), and generous support from private individuals and foundations.

The Institute for Gravitation and the Cosmos is supported by the Eberly College of Science and the Office of the Senior Vice President for Research at the Pennsylvania State University. The authors acknowledge the Texas Advanced Computing Center (TACC) at The University of Texas at Austin for providing high performance computing, visualization, and storage resources that have contributed to the research results reported within this paper. URL: \texttt{http://www.tacc.utexas.edu}.

\facility{HET, HST (WFC3)}

\software{Astropy \citep{astropy:2018}, NumPy \citep{numpy}, SciPy \citep{SciPy}, Matplotlib \citep{matplotlib}}


\clearpage
\bibliography{lya.bib}


\clearpage
\appendix

\section{Appendix: Recovering Co-added \lya Fluxes}
\label{sec:appendix}

To test our ability to measure \lya from stacked VIRUS spectra, we modeled the effects that the grism redshift uncertainty, the systematic velocity offset between the \lya and rest-frame optical emission lines, the VIRUS astrometric uncertainties, and the spectrum smoothing kernel have on our ability to measure co-added \lya line fluxes.  To simulate these effects, we constructed a set of model VIRUS spectra, applied the observational effects, stacked the data, and compared the recovered line fluxes to those of the input model.  To isolate the effect of the smoothing kernel we performed our experiment with and without spectral smoothing. We also repeated the experiment for two different line fluxes that reflect the range of line strengths displayed in Figure~\ref{fig:fits}.

The first step in simulating a galaxy was to mimic the astrometric precision of the VIRUS measurements.  A positional error in a VIRUS aperture affects the weights of the optimal extraction algorithm, and can lead to an error in an object's summed 1-D spectrum.  We simulated this effect by creating model and recovered extraction weights, where the latter are constructed by perturbing the true ($x,y$) positions of each object by offsets consistent with the typical astrometric uncertainty of a HETDEX pointing ($\sim 0\farcs 2$).

The central wavelength of a galaxy's \lya emission is determined by two effects: the redshift uncertainty produced by the low resolution of the WFC3's G141 grism and the offset between a galaxy's systematic velocity and its \lya emission.   We simulated these effects by first assigning a galaxy's systemic redshift using a uniform distribution spanning the redshift interval of our data ($1.90 < z < 2.35$), and then offsetting the \lya central wavelength from this redshift with a Gaussian deviate centered at $\Delta v = 350$~km~s$^{-1}$ with a standard deviation of 150~km~s$^{-1}$. (This perturbation is chosen to approximately match the \lya velocity offsets seen by \citet{Hashimoto_2013}, \citet{Shibuya_2014}, \citet{Trainor_2015}, and \citet{Muzahid_2020}.)  The error in the \threedhst\ redshift is then simulated by applying an additional shift to \lya using a value consistent with the $\sigma_z \sim 0.002 \times (1+z)$ grism redshift uncertainty. 


The galaxy's \lya profile is modeled as a Gaussian centered at the observed \lya wavelength computed above, \mbox{$\lambda_{\rm obs} = (1 + z_{{\rm Ly}\alpha}) \times 1216$~\AA}, with a width that matches the spectral resolution of the VIRUS instrument (FWHM $\sim 5.6$~\AA\null).  The amplitude of this Gaussian is set such that the integrated flux of the line matches the flux measured from the stacked spectra (e.g., the examples shown in Figure~\ref{fig:fits}). We repeated the experiment for two different flux levels, $f_{\rm Ly \alpha} = 2$~and~$4 \times 10^{-17}$~\ecs. 

\begin{figure*}
    \centering
    \includegraphics[width=0.475\linewidth]{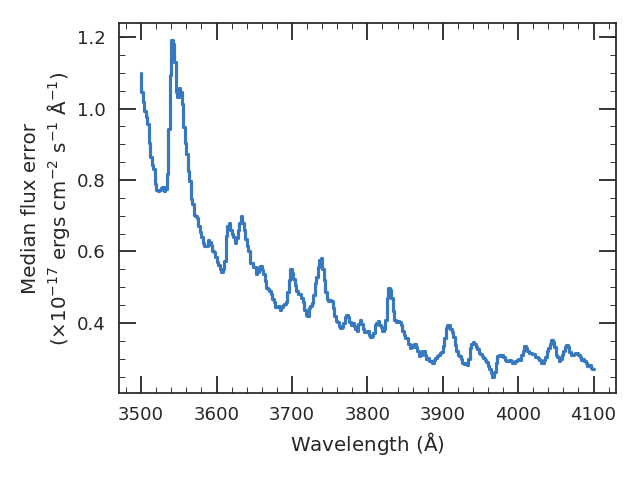}
    \caption{ The median flux errors for the HETDEX science verification observations of the \nsample\ objects in our sample. This curve is used to perturb the model spectra in the flux-recovery simulations. We restrict the wavelength range to that where \lya is observed in our $1.90<z<2.35$ sample.}
    \label{fig:median-flux-error}
\end{figure*}

After modeling each galaxy's VIRUS spectrum, our ``observation'' is performed by adding in the median noise (per pixel) of the VIRUS dataset (see Figure~\ref{fig:median-flux-error}), extracting the spectrum using the weights that incorporate the astrometric uncertainties, (optionally) smoothing the spectrum by 670~km~s$^{-1}$, and shifting the model to the rest-frame using the assumed \threedhst\ redshift.  

\begin{figure*}
  \captionsetup[subfigure]{labelformat=empty}
  \centering
  \subfloat[][]{\label{}%
    \includegraphics[width=0.475\linewidth]{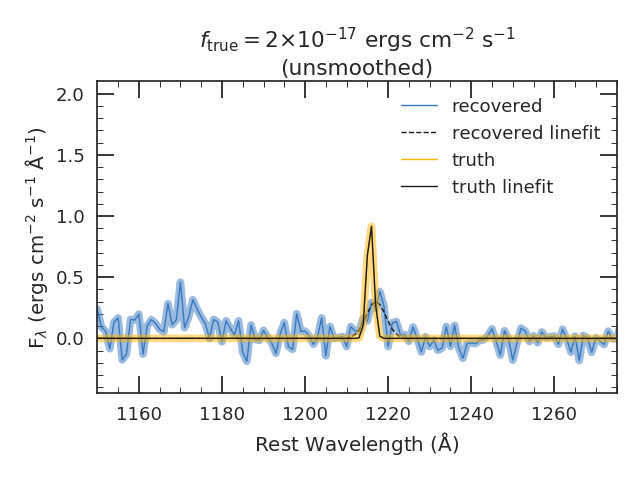}%
  }\hfill
  \subfloat[][]{\label{}%
    \includegraphics[width=0.475\linewidth]{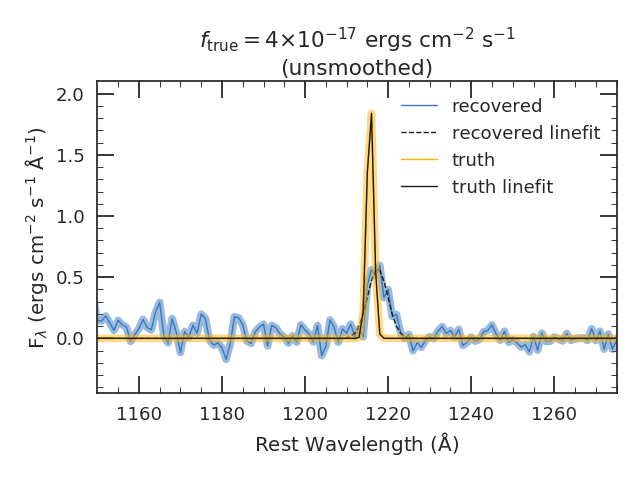}%
  }
  \vspace{-10 mm}
  \subfloat[][]{\label{}%
    \includegraphics[width=0.475\linewidth]{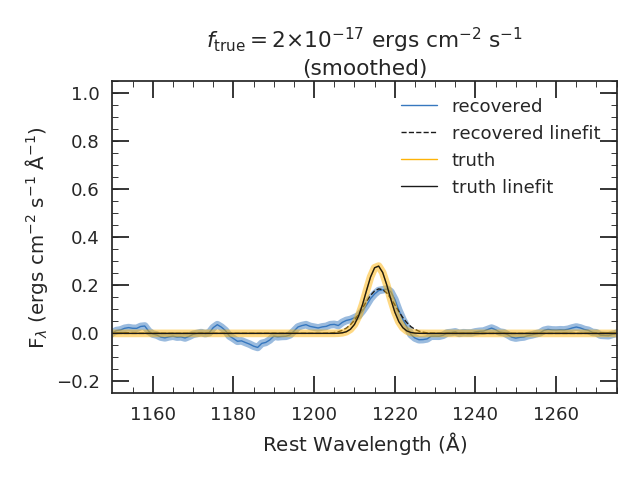}%
  }\hfill
  \subfloat[][]{\label{}%
    \includegraphics[width=0.475\linewidth]{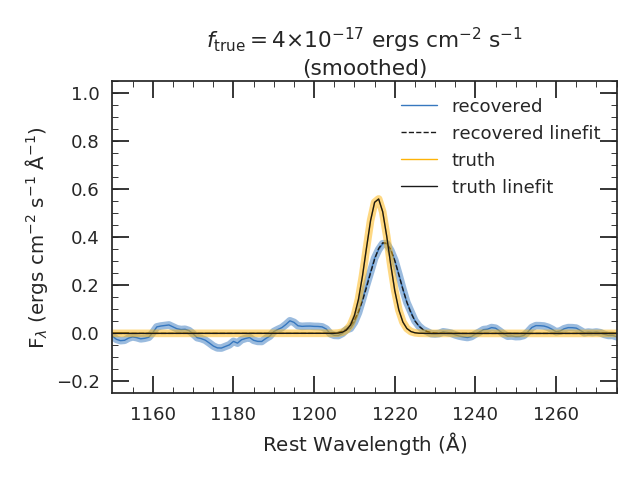}%
  }
  \caption{ The model and ``observed'' stacked spectra (realized from 200 individual spectra) for the two line fluxes used in the simulation, with and without the 670~km~s$^{-1}$ smoothing.}
  \label{fig:simulated-line-profiles}
\end{figure*}

This procedure was repeated 200 times, matching the typical number of galaxies that go into each stack described in this study. The final true and recovered line fluxes were then measured using the biweight of the stacked spectra. One instance of the stacked spectra (for each line flux level with and without the smoothing) is shown in Figure~\ref{fig:simulated-line-profiles}.

\begin{figure*}
  \captionsetup[subfigure]{labelformat=empty}
  \centering
  \subfloat[][]{\label{}%
    \includegraphics[width=0.45\linewidth]{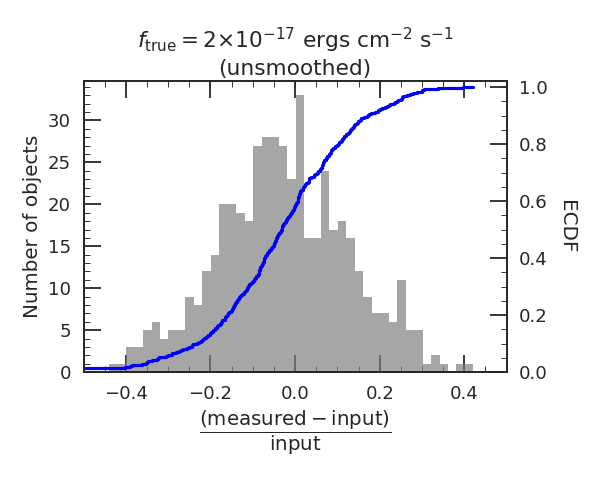}%
  }\hfill
  \subfloat[][]{\label{}%
    \includegraphics[width=0.45\linewidth]{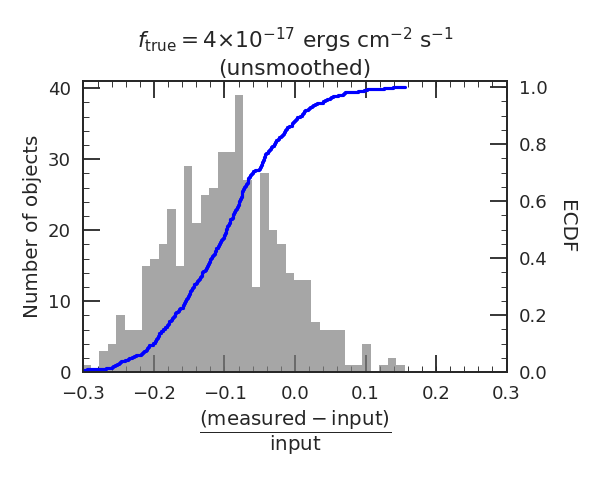}%
  }
  \vspace{-7.5 mm}
  \subfloat[][]{\label{}%
    \includegraphics[width=0.45\linewidth]{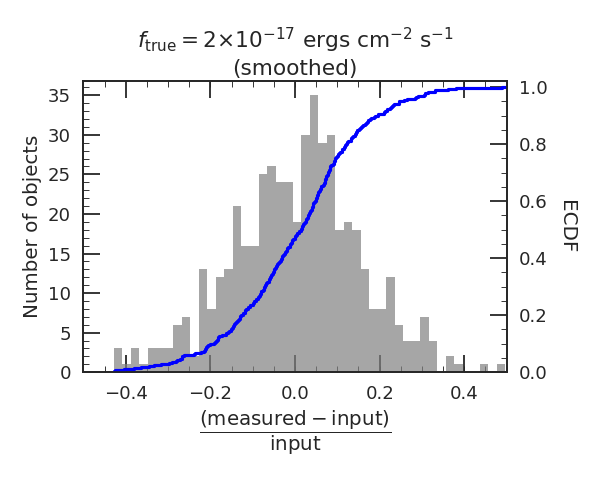}%
  }\hfill
  \subfloat[][]{\label{}%
    \includegraphics[width=0.45\linewidth]{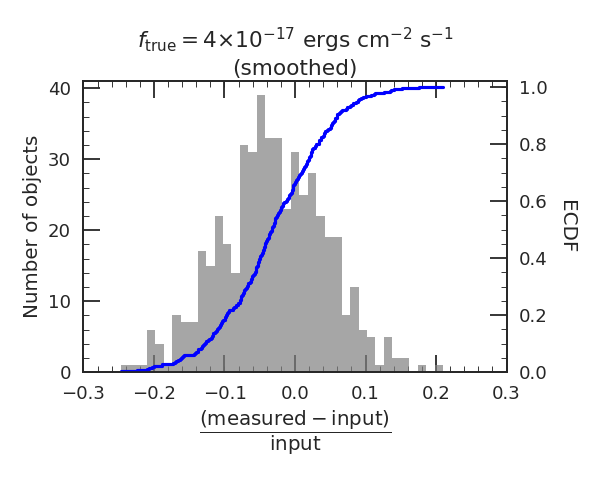}%
  }
  \caption{ The fractional offset between the model and recovered line fluxes measured from the stacked spectra. The experiment is repeated 500 times at each of the model line fluxes, with and without the 670~km~s$^{-1}$ spectral smoothing. When no smoothing is applied, there is a tendency to miss some of the line flux (i.e., the recovered values are systematically lower than the input line fluxes). No such bias is present when the spectra are smoothed prior to stacking, and the recovered values are typically within $\sim 0.3 \times 10^{-17}$~\ecs\ of the true line flux.
  }
  \label{fig:simulated-flux-recovery}
\end{figure*}

Figure~\ref{fig:simulated-flux-recovery} compares our model and recovered \lya fluxes for 500 simulated stacks.  When the line profiles are not smoothed by the 670~km~s$^{-1}$ kernel, the measurements from the stacked spectra tend to underestimate input fluxes by $\sim 10\%$, particularly in the higher flux simulations.  When the smoothing is applied, no such bias is present, and the recovered line fluxes scatter symmetrically about the true value with a $1\,\sigma$ dispersion of $\Delta f \sim 0.3 \times 10^{-17}$~\ecs.


\end{document}